\documentclass[12pt]{article}%
\usepackage{graphics}
\renewcommand{\theequation}{\arabic{section}.\arabic{equation}}%
\begin{document}%
\def\d{{\mathrm{d}}}
\def\g{{\mbox{\sl g}}}%
\def\Box{\nabla^2}%
\def\d{{\mathrm d}}%
\def\R{{\rm I\!R}}%
\def\ie{{\em i.e.\/}}%
\def\eg{{\em e.g.\/}}%
\def\etc{{\em etc.\/}}%
\def\etal{{\em et al.\/}}%
\markright{Causal structure of acoustic spacetimes\hfil}%
\title{\bf \LARGE Causal structure of acoustic spacetimes}%
\author{Carlos Barcel\'o~$^*$, Stefano Liberati~$^\dagger$,
Sebastiano Sonego~$^\ddagger$, and Matt Visser~$^\S$%
\\[2mm]%
{\small\it%
\thanks{\tt carlos@iaa.es}%
\ Instituto de Astrof\'{\i}sica de Andaluc\'{\i}a (CSIC),
Camino Bajo de Hu\'etor 24, 18008 Granada, Spain}%
\\[4mm]%
{\small\it%
\thanks{\tt liberati@sissa.it; http://www.sissa.it/\~{}liberati}%
\ International School for Advanced Studies (SISSA),
Via Beirut 2-4, 34013 Trieste, Italy}
\\ {\small \it%
and INFN, Trieste, Italy}%
\\[4mm]%
{\small\it%
\thanks{\tt sebastiano.sonego@uniud.it}%
\ Universit\`a di Udine, Via delle Scienze 208, 33100 Udine, Italy}%
\\[4mm]%
{\small\it%
\thanks{\tt matt.visser@mcs.vuw.ac.nz; http://www.mcs.vuw.ac.nz/\~{}visser}
\ School of Mathematical and Computing Sciences,
Victoria University of Wellington, New Zealand}}%
\date{{\small 8 August 2004;  V2: 1 Sept 2004; {\LaTeX-ed \today}; gr-qc/0408022}}%
\font\hn = phvrrn
\maketitle%
\begin{abstract}%
  
  The so-called ``analogue models of general relativity'' provide a
  number of specific physical systems, well outside the traditional
  realm of general relativity, that nevertheless are well-described by
  the differential geometry of curved spacetime. Specifically, the
  propagation of acoustic disturbances in moving fluids are described
  by ``effective metrics'' that carry with them notions of ``causal
  structure'' as determined by an exchange of sound signals.  These
  acoustic causal structures serve as specific examples of what can be
  done in the presence of a Lorentzian metric without having recourse
  to the Einstein equations of general relativity. (After all, the
  underlying fluid mechanics is governed by the equations of
  traditional hydrodynamics, not by the Einstein equations.) In this
  article we take a careful look at what can be said about the causal
  structure of acoustic spacetimes, focusing on those containing sonic
  points or horizons, both with a view to seeing what is different
  from standard general relativity, and to seeing what the
  similarities might be. %

\vspace*{5mm}%
\end{abstract}%

\clearpage
\section{Introduction}%
\label{sec:ac-spacetimes}%
\setcounter{equation}{0}%

Acoustic spacetimes are curved Lorentzian manifolds that are used to
describe the propagation of sound in moving
fluids~\cite{Unruh1,Jacobson,Visser1,Unruh2,visser98,abh}. As such
they are equipped with a Lorentzian spacetime metric (strictly
speaking, a pseudo-metric) that is associated with the ``sound cones''
emanating from each event in the spacetime. Though the acoustic
spacetimes are in general curved manifolds, with nonzero Riemann
tensor, their curvature and evolution (their geometrodynamics) are not
determined by the Einstein equations, but are instead implicit in the
equations of traditional fluid mechanics~\cite{Visser1,visser98,abh}.
The acoustic effective geometry must not be confused with the ``real''
spacetime geometry. Indeed, the ``real'' physical spacetime structure
that we experience in a condensed matter laboratory is approximately
Minkowskian.  Moreover, under normal circumstances the velocities
involved are so small in comparison with the velocity of light that we
can perfectly well assume that the system is non-relativistic
(Galilean), be it classical or quantum.%

Because of these features, the acoustic spacetimes play a rather
special role with respect to traditional general relativity.  They are
examples of Lorentzian manifolds without ``gravity'' and their
existence forces us to think deeply and carefully about the
distinction between kinematics and dynamics in general relativity ---
specifically how much of standard general relativity depends on the
Einstein equations and how much of it depends on more general
considerations that continue to hold independently of the Einstein
equations? In particular, this forces us to think about the deep
connections and fundamental differences between Lorentzian geometry,
the Einstein equivalence principle, and full general relativity.  Some
features that one normally thinks of as intrinsically aspects of
gravity, both at the classical and semiclassical levels (such as
horizons and Hawking radiation), can in the context of acoustic
manifolds be instead seen to be rather generic features of curved
spacetimes and quantum field theory in curved spacetimes, that have
nothing to do with gravity \emph{per
se\/}~\cite{Unruh1,visser98,Visser3,essential}.%

In this article we will develop an entire bestiary of
(1+1)-dimensional acoustic geometries, specifically chosen because of
their naturalness from the point of view of flowing fluids. We will
focus on the particularly interesting cases in which the acoustic
geometries possess one or two sonic points. These geometries will be
the starting point of a follow-up paper in which we will investigate
their different effects in terms of curved-space quantum field
theoretic vacuum polarization.

After describing each of these geometries, we will investigate their
global causal structure by the use of Carter--Penrose diagrams.  These
diagrams make it clear that because the acoustic geometries are not
governed by the Einstein equations, their causal structure can be
quite different from what is usually encountered in the context of
general relativity. In this context, we will also discuss the notion
of ``maximal analytic extension'' in these acoustic geometries on both
mathematical and physical grounds.  While mathematically the notion of
analytic extension makes perfectly good sense, there are now good
physical reasons for being cautious. This \emph{may\/} have
implications for physical gravity and in particular for the ability to
characterize spacetime structure by a single well-behaved metric (as
opposed to a multi-metric theory). Since this is the first step
towards implementing any version of the equivalence principle, it
strikes at the very foundations of general relativity.

The geometrical analyses of acoustic spacetimes we are going to
present are also interesting for two additional reasons:
\begin{enumerate}%
\item Because we have a very specific and concrete physical picture
  for these acoustic spacetimes it is sometimes easier for a
  classically trained physicist to see what is going on, and to then
  use this as a starting point for investigations of the perhaps more
  formal causal structure in standard general relativity.
\item Conversely, relativists can adopt their training to ask
  questions in acoustics that might not normally occur to classically
  trained acoustic physicists.
\end{enumerate}%

After dealing with some fundamental issues in section
\ref{S:fundamental} we shall introduce the concept of null coordinates
in section \ref{sec:null}. In section \ref{sec:stat-zoo} we develop a
``zoo'' of stationary acoustic spacetimes, focussing on situations
with either one or two sonic points.  Section \ref{sec:coll-zoo} looks
at the dynamical evolution of acoustic horizons as the fluid flow is
switched on from zero flow to the fully developed flows considered in
section \ref{sec:stat-zoo}. In sections \ref{sec:compact-stat} and
\ref{sec:compact-dyn} we investigate the global causal structure of
the stationary and dynamical acoustic spacetimes by the use of
Penrose--Carter diagrams, while the mathematical possibility of
performing an analytical extension for acoustic spacetimes will be
considered and discussed in section \ref{sec:regular}.  Finally, our
summary and conclusions are presented in section \ref{sec:comments}.

\section{Fundamental features}%
\label{S:fundamental}%
\setcounter{equation}{0}%

We start by pointing out that in acoustic spacetimes, as in general
relativity, causal structure can be characterized in two complementary
ways --- in terms of the rays of geometrical acoustics/optics or in
terms of the characteristics of the partial differential equations
(wave equations) of physical acoustics/optics~\cite{abh,normal}.

At the level of geometrical acoustics we need only assume that:
\begin{itemize}%
\item the speed of sound $c$, relative to the fluid, is well
defined;%
\item the velocity of the fluid $\vec v$, relative to the
laboratory, is well defined.%
\end{itemize}%
Then, relative to the laboratory, the velocity of a sound ray
propagating, with respect to the fluid, along the direction
defined by the unit vector $\vec n$ is%
\begin{equation}%
\frac{\d\vec{x}}{\d t}=c\,\vec{n}+\vec{v}\;,%
\end{equation}%
which defines a sound cone in spacetime given by the condition
$\vec{n}^2=1$, i.e.,%
\begin{equation}%
- c^2 \d t^2 + \left(\d\vec{x} - \vec v \,\d t \right)^2 = 0\;.%
\end{equation}%
This is associated with a conformal class of Lorentzian
metrics~\cite{visser98,abh}%
\begin{equation}%
\g  = \Omega^2\left[ \begin{array}{c|c}
-(c^2-v^2) & -{{\vec v}\,}^T\\
\hline -{\vec v} & {\bf I}
\end{array}\right]\;,%
\label{g-matrix}%
\end{equation}%
where $\Omega$ is an unspecified non-vanishing function.
The virtues of the geometric approach are its extreme
simplicity and the fact that the basic structure is
dimension-independent.%

At the level of physical acoustics, setting up an acoustic
spacetime is a little trickier. For technical reasons it is
easiest to confine attention to an irrotational flow for a
fluid with a barotropic equation of state, in which case it is
relatively straightforward to derive, in any number of dimensions,
a wave equation of the form~\cite{visser98,abh}%
\begin{equation}%
\partial_a \left( f^{ab}\, \partial_b \theta \right) = 0\;.%
\label{unlabeled}%
\end{equation}%
Turning this into a statement about a metric requires the
identification%
\begin{equation}%
\left(-\det \g\right)^{1/2} \, \g^{ab} = f^{ab}\;,%
\label{identification}
\end{equation}%
where, as usual, $\g$ denotes the matrix $[\g_{ab}]$,
obtained by inverting $\g^{-1}\equiv [\g^{ab}]$.  Defining
now $\det f = 1/\det[f^{ab}]$ we have%
\begin{equation}%
-\det \g = (- \det f)^{-2/(d-2)}\;,
\end{equation}%
and%
\begin{equation}%
\g^{ab} = (- \det f)^{1/(d-2)} \, f^{ab}\;.%
\label{problem}%
\end{equation}%
With this procedure one gets, for a fluid with mass density
$\rho$, a metric of the form (\ref{g-matrix}), with $\Omega^2$
equal to an unspecified positive constant multiplied by 
some power of $\rho/c$.

However, the exponent in equation (\ref{problem}) indicates that
$\g^{ab}$ and $\g_{ab}$ are not defined in $d=2$ (that is, in 1+1
dimensions). Fortunately this problem is formal rather than
fundamental.  One can always augment any interesting (1+1)-dimensional
acoustic geometry by two extra flat space dimensions --- which is
after all exactly how such a geometry would actually be experimentally
realised, letting the fluid flow along a thin pipe --- and simply
phrase physical questions in terms of the plane symmetric 3+1
geometry. Alternatively, one could forget the extra dimensions and
simply ask questions based on the geometric acoustics approximation,
which, after all, is quite sufficient for dealing with issues of
causal structure.

We mention in passing that attempts to include vorticity into the
physical acoustics formalism lead to a more complicated mathematical
structure of which the ``effective metric'' is only one part.
Fortunately the eikonal approximation (the geometrical acoustics
approximation) again leads to the conformal class of metrics
considered above and as far as issues of causal structure are
concerned, there is no significant gain in adding vorticity to the
mix~\cite{vorticity}.

Before going further we also mention that for any metric of the form
(\ref{g-matrix}) we have:%
\begin{equation}%
\g^{-1} = \frac{1}{\Omega^2} \left[ \begin{array}{c|c}
-1/c^2& -{{\vec v}\,}^T/c^2\\
\hline -{\vec v}/c^2 & {\bf I} - \vec v \otimes
{{\vec v}\,}^T/c^2
\end{array}\right]\;.%
\end{equation}%
Therefore, since%
\begin{equation}%
\g^{-1}(\d t, \d t) = - 1/(c^2\Omega^2) < 0\;,%
\end{equation}%
we see that the natural Newtonian time coordinate provides a
``cosmic time'' and so the acoustic manifolds are always stably
causal~\cite{visser98,abh}. This is one among many
special features exhibited by acoustic spacetimes.%

In the remainder of this paper we will always consider
(1+1)-dimensional acoustic spacetimes and will investigate their
causal structure in some detail.  From the laboratory point of view,
such a spacetime is always a Cartesian product between some open
interval $(t_1,t_2)$ of the real line (time) and some one-dimensional
manifold without boundary (space). We shall always consider these
spacetimes as eternally existing entities, so that in effect
$t_1=-\infty$ and $t_2=+\infty$.  It is well-known that any smooth,
connected one-dimensional manifold without boundary is diffeomorphic
either to $\R$ or to the circle $S^1$ \cite{milnor}.  Hence, the only
two possible diffeomorphism classes for our acoustic manifolds are
either ${\cal M}=\R^2$ or ${\cal M}=\R\times S^1$.  This simple
observation already greatly constrains the topology of our acoustic
spacetimes.  Additionally, on these manifolds there are preferred
coordinates, that we denote $t$ and $x$, associated with time and
distance readings by means of physical Newtonian clocks and rulers.
We have $t\in\R$, and $x\in\R$ or $x\in(-L,L)$, according to whether
space is topologically a line or a circle. Note that the acoustic
spacetimes, armed with such a notion of measurement, are automatically
inextensible manifolds for physical reasons.

As we have said before, associated with the irrotational flow of
a barotropic, viscosity-free fluid, there is an {\em acoustic
metric\/} on $\cal M$,%
\begin{equation}%
\g=\Omega^2\left[-\left(c^2-v^2\right)\d t^2
-2\,v\,\d t\,d x+\d x^2\right]\;,%
\label{metric}%
\end{equation}%
where $c$ is the speed of sound and $v$ the fluid velocity
\cite{Unruh1,Visser1,visser98,abh}.  Now in general, $c$ and $v$
are functions of $t$ and $x$. Here, we shall in the interests of
simplicity assume that $c$ does not depend on position and
time. It is therefore the velocity profile $v(t,x)$ that contains
all the relevant information about the causal structure of
the acoustic spacetime $({\cal M},\g)$.  %

The effective metric (\ref{metric}) is only experienced by acoustic
waves (and phonons) in the fluid.  Hypothetical ``internal''
observers, living in the fluid and able to measure distances and times
only by means of acoustic structures, might consider the proper
distances associated with this acoustic metric to be the actual
physical distances~\cite{visser98,paradox}, at least for low
energies~\cite{breaking}.  (An early suggestion somewhat along these
lines can be found in reference~\cite{nandi}.)  In contrast, from the
``external'' laboratory point of view proper distances in $({\cal
  M},\g)$ are meaningless, since it is only $t$ and $x$ that have
empirical content in terms of external laboratory measurements.

Topologically, the manifold $\cal M$ is an open set. As we have
already mentioned, from the laboratory point of view it is an
inextensible manifold. However, from the point of view of the internal
observers this manifold might be extensible for certain particular
acoustic configurations. Roughly speaking, this can happen when the
proper distance to infinity in the acoustic metric becomes finite. In
this paper we will analyze configurations with one or more sonic
points, in which these phenomena usually happen. These sonic points
are points where $v(t,x)=\pm c$, and correspond to the so-called
acoustic apparent horizons\footnote{We caution the reader that
the relative simplicity of the discussion regarding horizons and sonic
points depends to a large extent on the simplification of dealing with
1+1 dimensions. In higher dimensions without planar symmetry one
should generically distinguish horizons from
ergosurfaces~\cite{visser98}.} for the Lorentzian geometry defined on
$\cal M$ by the metric (\ref{metric}). In fact, we shall use the
terminology ``sonic point'' and ``apparent horizon''
interchangeably. At the sonic points, the coefficient of $\d t^2$
vanishes, whereas the determinant of the metric is everywhere equal to
$-c^2$, hence regular. This suggests that the coordinates $t$ and $x$
are somehow inappropriate to describe sound propagation near the sonic
points.%

Indeed, as we shall see, although by definition $t$ and $x$ cover
the whole manifold, the behaviour of null curves in a $(t,x)$
diagram can be singular at the horizons.  Moreover, $t$ and $x$
convey a wrong impression of the acoustic causal structure, since
they are not everywhere timelike and spacelike coordinates with
respect to the acoustic metric.%

Our primary goals in this article are to give a clear description of
the causal structure of the most natural acoustic spacetimes with
sonic horizons; and to emphasize the similarities with, and
differences from, the standard black-hole solutions in standard
general relativity.%

\section{Null coordinates}%
\label{sec:null}%
\setcounter{equation}{0}%

In order to explore the causal structure of $({\cal M},\g)$, let
us introduce retarded and advanced null coordinates as%
\begin{equation}%
\d u:=F(t,x)\;\left(\d t-\frac{\d x}{c+v(t,x)}\right)\;,%
\label{U}%
\end{equation}%
\begin{equation}%
\d w:=G(t,x)\;\left(\d t+\frac{\d x}{c-v(t,x)}\right)\;,%
\label{W}%
\end{equation}%
where $F$ and $G$ are suitable integrating factors. In these
coordinates, the metric is%
\begin{equation}%
\g=-\frac{\Omega^2}{F\,G}\left(c^2-v^2\right)\,\d u\,\d w\;.%
\label{metric-UW}%
\end{equation}%
Of course, for any stationary flow, in which $v$ does not depend
on $t$, we can choose $F\equiv G\equiv 1$.  Let us consider this
case in some detail.%

Consider a smooth velocity profile with a sonic point at
$x=x_S$.  Then one can write, for $x$ sufficiently close to $x_S$,%
\begin{equation}%
v(x)=\sigma c +\epsilon \kappa\,\left(x-x_S\right) +O\left(
\left(x-x_S\right)^2 \right)\;,%
\label{v}%
\end{equation}%
where $\sigma$ and $\epsilon$ can take on both the values
$\pm 1$, and%
\begin{equation}%
\kappa:=\left|\frac{\d v}{\d x}\right|_{x=x_S}%
\label{def:kappa}%
\end{equation}%
is a nonnegative quantity which can be interpreted as the normalized
surface gravity~\cite{visser98,essential,surface-g}. There are now
four possibilities depending on the values of $\sigma$ and $\epsilon$:
\begin{enumerate}%
\item[(i)] At the sonic point $\sigma=-1$ (left-going flow)
and $\epsilon=1$; this corresponds to a black-hole configuration
from the point of view of an observer to the right of $x_S$;%
\item[(ii)] At the sonic point $\sigma=1$ (right-going flow)
and $\epsilon=-1$; to an observer to the right of $x_S$,
this corresponds to a white-hole configuration;%
\item[(iii)] and (iv) correspond, respectively, to mirror symmetric
configurations of {(i)} and {(ii)} through the point $x_S$
(right and left are exchanged).%
\end{enumerate}%

When $\sigma=-1$, $w$ is regular at $x=x_S$, while $u$
diverges.  The asymptotic expressions near the sonic point are%
\begin{equation}%
\left.\begin{array}{l}%
{\displaystyle u\sim t-\frac{1}{\kappa}\,\ln|x-x_S|}\\%
\\%
{\displaystyle w\sim t+\frac{x}{2\,c}}%
\end{array}\right\}\;.%
\label{U-asympt}%
\end{equation}%
When $\sigma=1$, it is $u$ that is regular, while $w$
diverges.  The asymptotic expressions are%
\begin{equation}%
\left.\begin{array}{l}%
{\displaystyle u\sim t-\frac{x}{2\,c}}\\%
\\%
{\displaystyle w\sim t + \frac{1}{\kappa}\,\ln|x-x_S|}%
\end{array}\right\}\;.%
\label{W-asympt}%
\end{equation}%
In both cases, the equation $x=x_S$ defines a null line in
spacetime, where either $u$ or $w$ tend to $+\infty$.  Obviously,
the null coordinates $u$ and $w$ are unsuitable to describe the
geometry in the vicinity of the sonic points, as one can see also
from the corresponding expression (\ref{metric-UW}) of the
metric.%

For completeness, we mention that the laboratory coordinates
$t$ and $x$ are not the analogs of the Schwarzschild-like
coordinates commonly used when investigating static solutions of
Einstein's equations with spherical symmetry.  This is obvious if
one thinks (i) that the lines $t=\mbox{const}$ are not orthogonal,
in the acoustic metric (\ref{metric}), to the worldlines of static
observers located at $x=\mbox{const}$, and (ii) that $(t,x)$ are
well defined across the sonic points, whereas the Schwarzschild
coordinate $t$ is singular on the horizons.  In fact, the
coordinates $t$ and $x$ correspond instead to the so-called
Painlev\'e-Gullstrand coordinates \cite{pg}; see also reference
\cite{visser98}. In order to make this point totally clear, let us
introduce Schwarzschild-like coordinates on the relevant portion of
the acoustic spacetime.%

Starting from the null coordinates $u$ and $w$ given by equations
(\ref{U}) and (\ref{W}), define $\tau$ and $\chi$ such that:%
\begin{equation}%
\d\tau=\frac{1}{2}\,\left(\d w+\d u\right)=\d
t+\frac{v}{c^2-v^2}\,\d x\;;%
\label{t}%
\end{equation}%
\begin{equation}%
\d\chi=\frac{1}{2}\,\left(\d w-\d u\right)=
\frac{c}{c^2-v^2}\,\d x\;.%
\label{x}%
\end{equation}%
On replacing these expressions into the metric (\ref{metric-UW})
we find%
\begin{equation}%
\g=\frac{\Omega^2}{FG}\left(c^2-v^2\right)\left(-\d\tau^2+\d \chi^2\right)
=\frac{\Omega^2}{FG}
\left[-\left(c^2-v^2\right)\d\tau^2+\frac{c^2}{c^2-v^2}\,\d x^2\right]\;.%
\label{metric-tx}%
\end{equation}%
It is evident that $\tau$ and $x$ are now perfect analogs of the
Schwarzschild time and radial coordinates.  In particular, (i) the
lines $\tau=\mbox{const}$ are orthogonal to the worldlines of static
observers, and (ii) the coordinate $\tau $ fails to be defined at the
sonic points. On the other hand, the coordinate $\chi$ is the analogue
of the radial coordinate used in the so-called optical
geometry~\cite{optical}. It represents spatial distances as measured
in terms of the round-trip time of acoustic signals.%

\section{Stationary zoo}%
\label{sec:stat-zoo}%
\setcounter{equation}{0}%

In this section we will describe a series of
stationary\footnote{We are using ``stationary'' in the precise
technical sense of assuming the existence of a Killing vector that
is timelike in the domain of outer communication.}
(1+1)-dimensional spacetime geometries that naturally appear in
the context of moving fluids.  For each of these geometries we
will exhibit a representative and simple to manipulate velocity
profile $v(x)$.  Then, we will show the behaviour of right-
and left-moving sound rays in these manifolds by drawing
lines $u=\mbox{const}$ and $w=\mbox{const}$ in a $(t,x)$ diagram
(Painlev\'e-Gullstrand coordinates).  We will postpone to section
\ref{sec:compact-stat} the detailed description of the conformal
structures of these spacetimes, through the inspection of
their conformal diagrams.  Furthermore, we leave it for section
\ref{sec:regular} to show that their metric manifold $({\cal
M},\g)$ can be mathematically extended beyond the physical range of
the coordinates $(t,x)$, and to discuss the physical interpretation of
this ``maximal analytic extension''. 
We have also investigated general flows without sonic points,
and report that their causal structure (though not necessarily their
geometry) is trivial in the sense that the null geodesics (and indeed
the Penrose--Carter diagrams) are qualitatively similar to those of 
ordinary flat Minkowski space.

\subsection{Flows with one sonic point}%

Let us start this catalog of configurations by considering those that
possess a single sonic point.  Since we are at this stage considering
stationary spacetimes, the apparent and absolute event horizons will
always coincide. Unfortunately the term ``event horizon'' has now
become so overused that it has to a large extent ``lost its
trademark'' and quite often when one encounters discussion of ``event
horizons'' in the literature the phrase is being misused to vaguely
refer to some sort of unspecified horizon-like object.  We shall try
to be more careful and stick to the proper technical meaning.
Hereafter, we shall always use the term ``event horizon'' as a synonym
of the ``absolute horizon'', that is the boundary of the causal past
of a suitable connected component of future null infinity.  In 1+1
dimensions without periodic boundary conditions on space there is a
definite and unambiguous notion of ``left'' and ``right'' so that null
infinity (and indeed spacelike infinity) naturally subdivides into at
least two disconnected components. The definition of an event horizon
is then subordinated to first choosing the disconnected component of
null infinity that is of interest. Similar behaviour occurs, for
instance, in the maximally extended Reissner--Nordstr\"om or Kerr
geometries.  Whenever it is not mentioned explicitly, we always use
the right null infinity ${\Im}_\mathrm{right}$ in order to define
horizons and black (or white) hole regions.  That is, we always make
reference to observations performed in the asymptotic region $x\to
+\infty$.

\subsubsection{Black hole}%
\label{subsubsec:bh}%

\begin{figure}[htbp]
\vbox{
\vskip10pt
\centerline{
\scalebox{0.600}{{\includegraphics{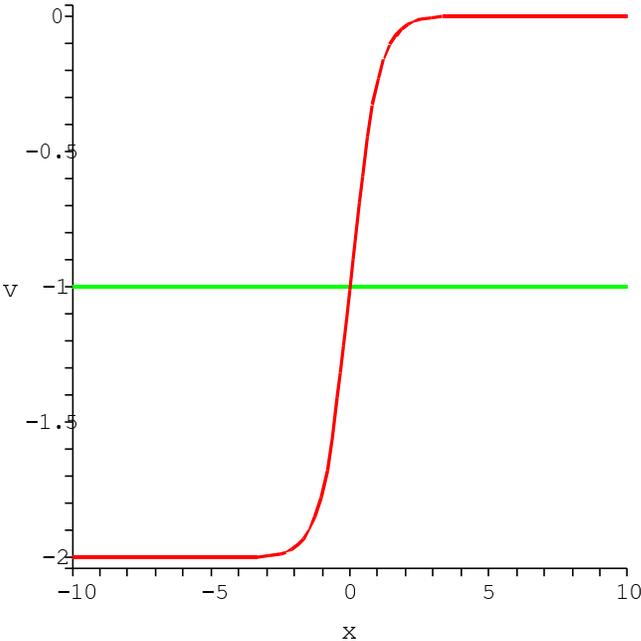}}}
}

\bigskip
\caption{
Velocity profile for a left-going flow with one subsonic region
($x>0$) and one supersonic region ($x<0$); $v$ bounded:  An
acoustic black hole.%
}
\label{F:v-bh}
}
\end{figure}

Consider a left-going flow, with one subsonic region and one
supersonic; with the flow velocity $v$ being bounded.  A velocity
profile that is, at the same time, both representative and simple to
manipulate is%
\begin{equation}%
v(x)=-\frac{2\,c}{\exp\left(2\,x/a\right)+1}\;,%
\label{v1}%
\end{equation}%
where $a$ is a positive constant.  This is plotted in
figure~\ref{F:v-bh} for $c=1$, $a=1$ (a choice of parameters
that we shall adopt for all diagrams in the paper).%

\noindent
Right-moving and left-moving sound rays are described by
$u=\mbox{const}$ and $w=\mbox{const}$, respectively, where:%
\begin{equation}%
u=t-\frac{x}{c}-\frac{a}{c}\ln\left|1
-\exp\left(-2\,x/a\right)\right|\;;%
\label{U1}%
\end{equation}%
\begin{equation}%
w=t+\frac{x}{c}+\frac{a}{3\,c}\ln\left(1
+3\,\exp\left(-2\,x/a\right)\right)\;.%
\label{W1}%
\end{equation}%
%
\begin{figure}[htbp]
\vbox{
\hfil
\scalebox{0.600}{{\includegraphics{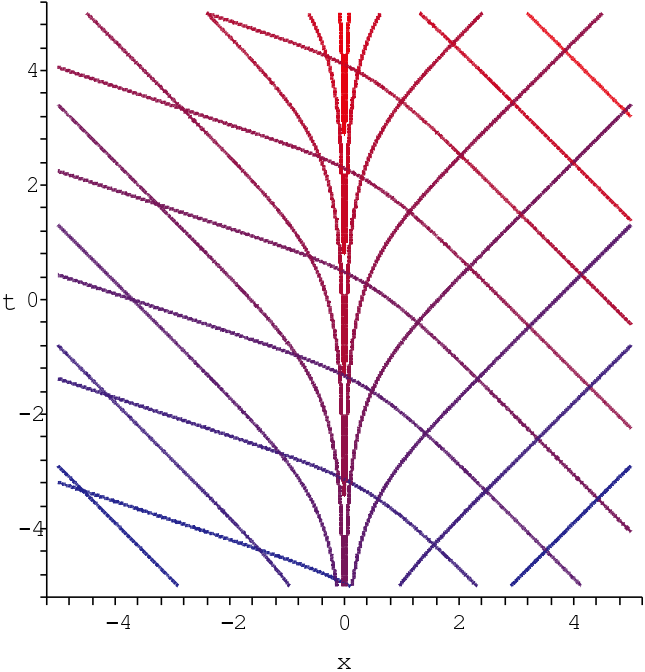}}}
\hfil
}
\bigskip
\caption{
Acoustic black hole. The null curves $u=\mathrm{const}$ are the
blue lines; $w=\mathrm{const}$ are the red lines.%
}
\label{F:UW-bh}
\end{figure}

\noindent Figure~\ref{F:UW-bh} represents the acoustic spacetime
with a few sound rays explicitly shown. The sound rays define the
acoustic causal structure. The acoustic horizon is located at
$x=0$ and the (normalized) surface gravity, defined as we have
already mentioned via $\kappa = \left|\d v/\d x\right|$ evaluated
at the apparent horizon, is $\kappa=c/a$.%

\subsubsection{White hole}%

\begin{figure}[htbp]
\vbox{
\hfil
\scalebox{0.600}{{\includegraphics{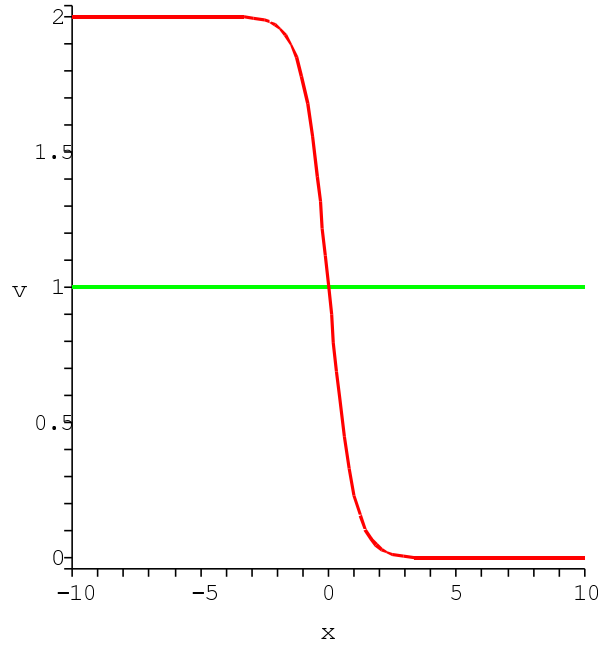}}}
\hfil
}
\bigskip
\caption{
Velocity profile for a right-going flow with one subsonic region
and one supersonic region; $v$ bounded:  An acoustic white
hole.%
}
\label{F:v-wh}
\end{figure}

\begin{figure}[htbp]
\vbox{
\hfil
\scalebox{0.600}{{\includegraphics{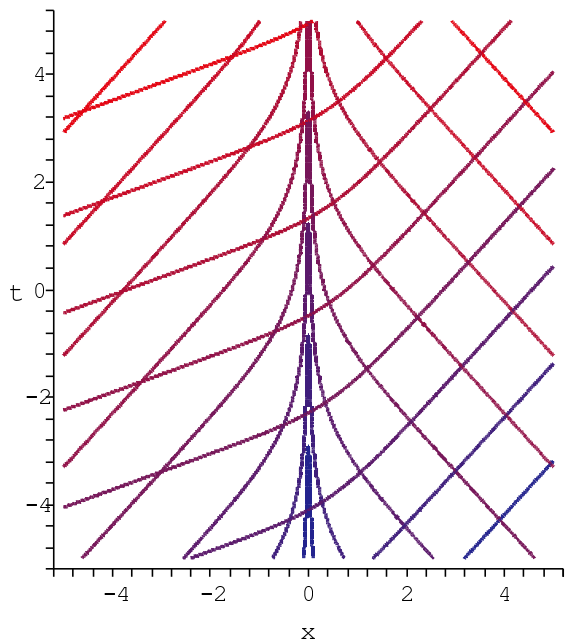}}}
\hfil
}
\bigskip
\caption{
Acoustic white hole. $u=\mathrm{const}$ are the blue lines,
$w=\mathrm{const}$ are the red lines.%
}
\label{F:UW-wh}
\end{figure}

Consider now a right-going flow, with one subsonic region, and one
supersonic; and with $v$ again bounded.  A convenient and easily
manipulated example is: %
\begin{equation}%
v(x)=\frac{2\,c}{\exp\left(2\,x/a\right)+1}%
\label{v2}%
\end{equation}%
(see figure~\ref{F:v-wh}).%
%
The null coordinates are%
\begin{equation}%
u=t-\frac{x}{c}-\frac{a}{3c}\ln\left(1
+3\,\exp\left(-2\,x/a\right)\right)\;,%
\label{U2}%
\end{equation}%
\begin{equation}%
w=t+\frac{x}{c}+\frac{a}{c}\ln\left|1
-\exp\left(-2\,x/a\right)\right|\;,%
\label{W2}%
\end{equation}%
and the acoustic spacetime is represented in figure~\ref{F:UW-wh}.
The acoustic horizon is again at $x=0$ and the surface gravity is
again $\kappa=c/a$.%
%

\subsubsection{Black hole, non-physical}%

\begin{figure}[htbp]
\vbox{
\hfil
\scalebox{0.600}{{\includegraphics{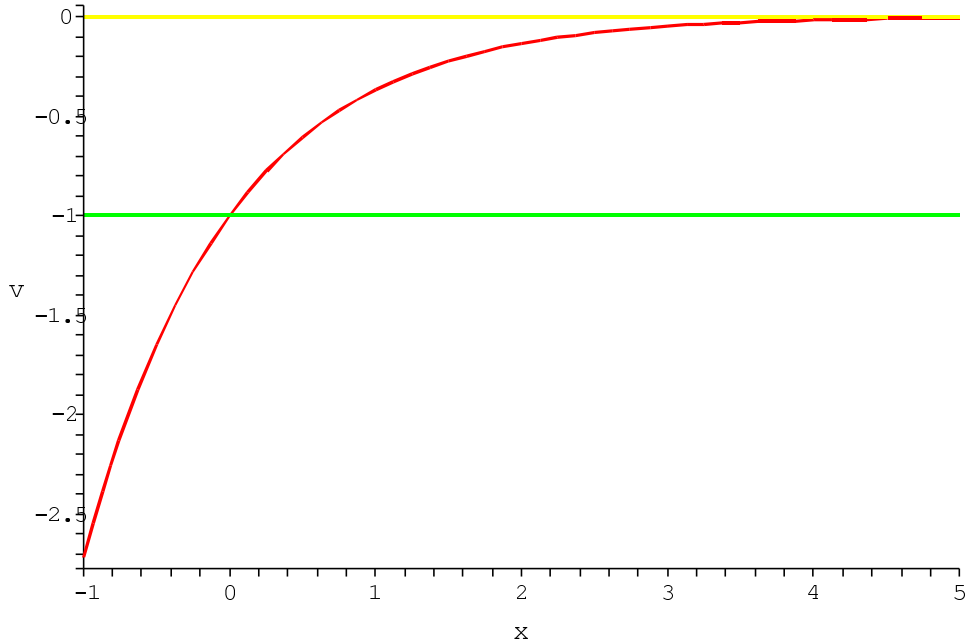}}}
\hfil
}
\bigskip
\caption{
Velocity profile for an unphysical black hole. There is again a
left-going flow, one subsonic region, one supersonic, but
now $v$ is unbounded.%
}
\label{F:v-unph-bh}
\end{figure}

\begin{figure}[htbp]
\vbox{
\hfil
\scalebox{0.600}{{\includegraphics{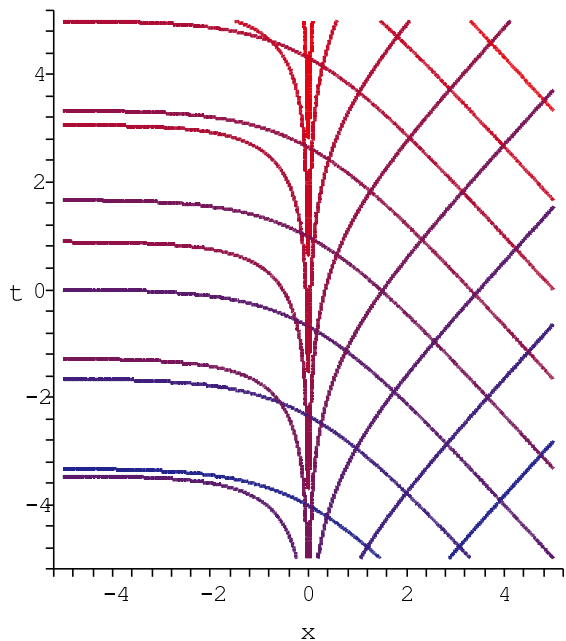}}}
\hfil
}
\bigskip
\caption{
Unphysical acoustic black hole. $u=\mathrm{const}$ are the blue
lines, $w=\mathrm{const}$ are the red lines.%
}
\label{F:UW-unph-bh}
\end{figure}

Consider a left-going flow, one subsonic region, one
supersonic; but with the flow velocity $v$ now unbounded. For
example, take%
\begin{equation}%
v(x)=-c\exp\left(-x/a\right)%
\label{v3}%
\end{equation}%
(as sketched in figure~\ref{F:v-unph-bh}), which leads to%
%
\begin{equation}%
u=t-\frac{x}{c}-\frac{a}{c}\ln\left|1
-\exp\left(-x/a\right)\right|\;,%
\label{U3}%
\end{equation}%
\begin{equation}%
w=t+\frac{x}{c}+\frac{a}{c}\ln\left(1
+\exp\left(-x/a\right)\right)\;,%
\label{W3}%
\end{equation}%
and to the acoustic spacetime of figure~\ref{F:UW-unph-bh}.  From
a laboratory point of view, this configuration is unphysical, as
the velocity of the flow grows without limit. Though unphysical
when interpreted in terms of a flowing fluid, we have nevertheless
included this example here because it is similar to what actually
happens in a Schwarzschild black hole when it is put into
``acoustic form''.  In Painlev\'e-Gullstrand coordinates the
velocity profile associated with a Schwarzschild black hole
($v_{\rm S} \propto 1/\sqrt{r}$; see \cite{visser98} for example)
grows without limit once the horizon is crossed. The divergence of
the velocity profile there signals the appearance of a
singularity.  Here, the same happens as $x\to -\infty$. Indeed,
the Ricci scalar (which, in two dimensions, contains all the
information about curvature) for the acoustic metric
(\ref{metric}) is, when $v$ does not depend on $t$,%
\begin{equation}%
R=\frac{1}{c^2\,\Omega^2}\,\frac{\d^2 v^2}{\d x^2}+\mbox{terms
containing derivatives of $\Omega$}\;.%
\end{equation}%
For the velocity profile (\ref{v3}) this becomes%
\begin{equation}%
R=\frac{4}{a^2\,\Omega^2}\,\exp\left(-2\,x/a\right)+\mbox{terms
containing derivatives of $\Omega$}\;,%
\end{equation}%
which, for a regular $\Omega$, diverges when $x \to -\infty$,
indicating the presence of a curvature singularity.  Indeed, all
of the standard curvature invariants diverge exponentially as
$x\to -\infty$.%

The difference with respect to the Schwarzschild black hole is
that there, the divergence appears at a finite coordinate
distance. However, in both cases the proper
distance\footnote{In fact, proper time, as the
coordinate $r$ (or $x$ in this particular case) changes its
spatial character to a timelike character beyond the horizon.}
from a point beyond the horizon to the singularity is finite. We
will discuss this point further in section \ref{sec:regular}.  For
the specific example given above, the horizon is again at $x=0$
and the surface gravity is again $\kappa=c/a$.%

\subsubsection{Extremal black hole}%
\begin{figure}[htbp]
\vbox{
\hfil
\scalebox{0.600}{{\includegraphics{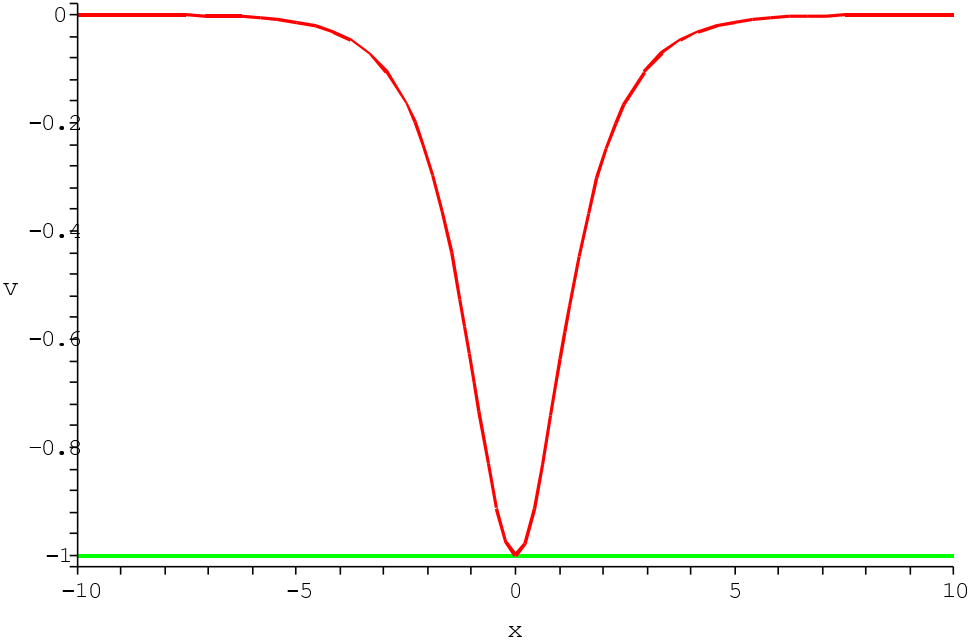}}}
\hfil
}
\bigskip
\caption{
Extremal black hole. Velocity profile for a
left-going flow, one sonic point, no supersonic region.%
}
\label{F:v-ex-bh}
\end{figure}

\label{staticextremal}%
Consider now a left-going flow, with one sonic point, and no
supersonic region:
\begin{equation}%
v(x)=-\frac{c}{\cosh\left(x/a\right)}%
\label{v5}%
\end{equation}%
(figure~\ref{F:v-ex-bh}).  For obvious reasons, we call this
acoustic geometry extremal because the derivative of the velocity
profile with respect to $x$ vanishes at $x_S$. Equivalently,
the parameter $\kappa$, representing the surface gravity at the
horizon, vanishes.  This is exactly what happens in a
Reissner--Nordstr\"om extremal geometry. The velocity profile (in
Painlev\'e--Gullstrand coordinates) for a Reissner--Nordstr\"om
extremal black hole is%
\begin{equation}%
v(r)^2=-{2M \over r}+ {M^2 \over r^2}.
\label{r-n}%
\end{equation}%
Once the horizon at $r_H=M$ is crossed, the velocity starts to
decrease until it becomes zero at $r=M/2$. The quantity $v(r)^2$
then changes sign and grows without limit, forming a singularity
at $r=0$. In a condensed matter laboratory this profile would be
unphysical, in particular since the velocity would be imaginary,
so the closest we can get is to make the velocity profile decrease
progressively towards zero once the horizon is crossed.
(Modifications of the Painlev\'e--Gullstrand coordinates for the
Reissner--Nordstr\"om geometry that avoid these imaginary
velocities have been discussed by Volovik~\cite{volovik}, see
also~\cite{heuristic}.)%

%
\begin{figure}[htbp]
\vbox{
\hfil
\scalebox{0.600}{{\includegraphics{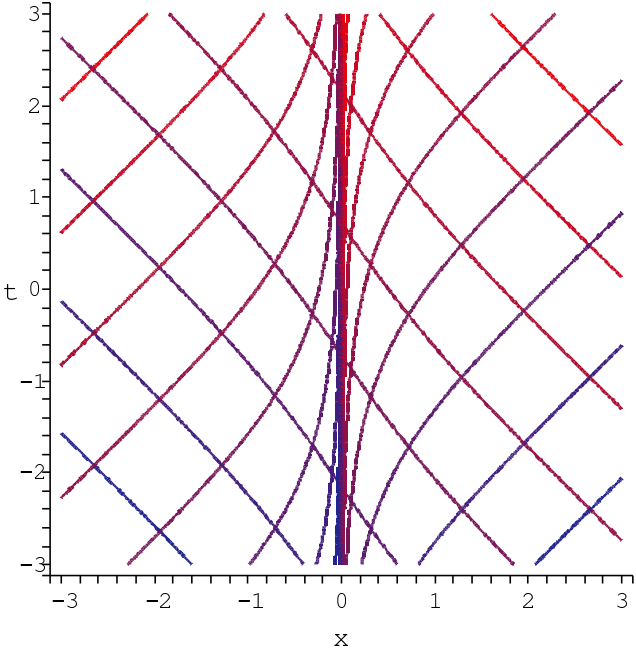}}}
\hfil
}
\bigskip
\caption{
Extremal black hole. $u=\mathrm{const}$ are the blue
lines, $w=\mathrm{const}$ are the red lines. This is qualitatively 
indistinguishable from a critical black hole spacetime. %
}
\label{F:UW-ex-bh}
\end{figure}

%
The null coordinates for this example are
\begin{equation}%
u=t-\frac{x}{c}+\frac{2\,a}{c\left(\exp\left(x/a\right)
-1\right)}\;;%
\label{U5}%
\end{equation}%
\begin{equation}%
w=t+\frac{x}{c}+\frac{2\,a}{c\left(\exp\left(x/a\right)
+1\right)}\;.%
\label{W5}%
\end{equation}%
The acoustic spacetime is represented in figure~\ref{F:UW-ex-bh}.
The horizon occurs at $x=0$ and by construction $\kappa=0$.%
%

\subsubsection{Critical black hole}%
\begin{figure}[htbp]
\vbox{
\hfil
\scalebox{0.600}{{\includegraphics{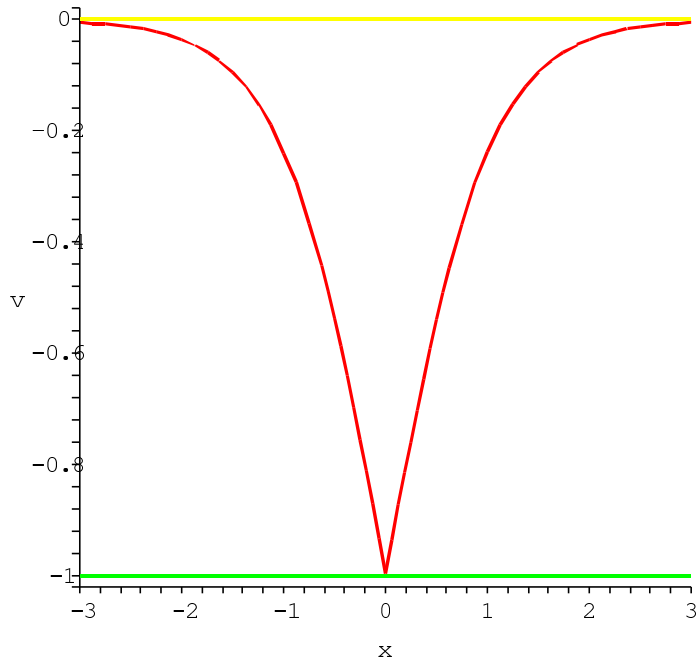}}}
\hfil
}
\bigskip
\caption{
Critical black hole: Velocity profile for a left-going flow, one sonic
point, no supersonic region, but singular flow at the sonic point.%
}
\label{F:v-spike-bh}
\end{figure}

Consider now a left-going flow, with one sonic point, no
supersonic region, but with a singular flow at the horizon. For
example%
\begin{equation}%
v(x) = - \frac{2 c}{\exp(2 |x|/a) + 1}%
\label{vspike}%
\end{equation}%
(figure~\ref{F:v-spike-bh}).  The major difference is now that,
because of the dependence on the absolute value of $x$,  the
velocity profile changes abruptly at the sonic point.%
%
%
%
The null coordinates for this example are:%
\begin{equation}%
u=\left\{\begin{array}{ll}
 t - {\displaystyle \frac{x}{c}
    - \frac{a}{c}\ln\left (\frac{\exp(-2 x/a)}{\exp(-2
      x/a) - 1}\right)} & \mbox{for $x<0$};\\
&\\
t - {\displaystyle \frac{x}{c}
    + \frac{a}{c}\ln\left(\frac{\exp(2 x/a)}{\exp(2 x/a) - 1}\right)}
& \mbox{for $x>0$};
\end{array}\right.
\label{Uspike}%
\end{equation}
\begin{equation}%
w=\left\{\begin{array}{ll} t +{\displaystyle  \frac{x}{c}
    - \frac{a}{3 c}\ln\left(\frac{4 \exp(-2 x/a)}{\exp(-2
      x/a)+3}\right)} & \mbox{for $x<0$};\\
&\\
t +{\displaystyle  \frac{x}{c}
    + \frac{a}{3 c}\ln\left(\frac{4 \exp(2 x/a)}{\exp(2
      x/a)+3}\right)} & \mbox{for $x>0$}.%
\end{array}\right.
\label{Wspike}%
\end{equation}%
This acoustic geometry is similar to that of the extremal case (see
figure~\ref{F:UW-ex-bh}) in that at the sonic point the velocity of
the flow does not change from subsonic to supersonic, but comes back
to subsonic. The important difference is that at the horizon the
one-sided surface gravity $\kappa$ is now different from zero.

Indeed the horizon is located at $x=0$, and if the surface gravity is
calculated in a one-sided manner from either side of the horizon we
find $\kappa=c/a$. However if the surface gravity is evaluated by
taking a two-sided symmetric limit then $\kappa=0$. This is a result
of the abrupt change in velocity profile at the horizon, so that this
particular model spacetime exhibits some features of both extremal and
non-extremal geometries.  This aspect of the model is particularly
interesting and might be important when analyzing the Hawking process
of quantum emission from the vacuum (we leave this analysis for a
follow-up paper).%

\subsection{Flows with two sonic points}%

Let us now consider the description of stationary but otherwise
arbitrary (1+1)-dimensional acoustic geometries in which there are
two sonic points.%

\subsubsection{Black and white hole combination}%

\begin{figure}[htbp]
\vbox{
\hfil
\scalebox{0.600}{{\includegraphics{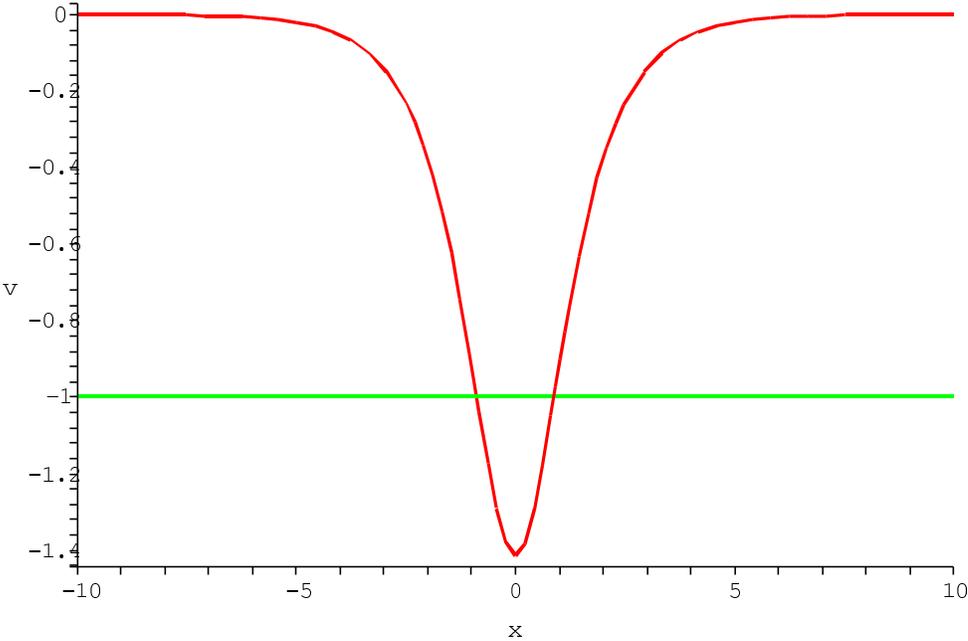}}}
\hfil
}
\bigskip
\caption{
Velocity profile for a black hole--white hole combination: A
left-going flow, two subsonic regions, one supersonic region
sandwiched between them; $\alpha=\sqrt{2}$.%
}
\label{F:v-bh-wh}%
\end{figure}

Such a combination can be realised by a left-going flow, with two
subsonic regions, and one supersonic region located between them.
Consider for example%
\begin{equation}%
v(x)=-\frac{c\,\alpha}{\cosh\left(x/a\right)}\;,%
\label{v4}%
\end{equation}%
with $\alpha>1$ (figure~\ref{F:v-bh-wh}).  The two sonic points
are $x_S=x_1$ and $x_S=x_2$, with%
\begin{equation}%
x_{1,2}=a\ln\left(\alpha\mp\sqrt{\alpha^2-1}\right)\;.%
\end{equation}%
%

%
\begin{figure}[htbp]
\vbox{
\hfil
\scalebox{0.600}{{\includegraphics{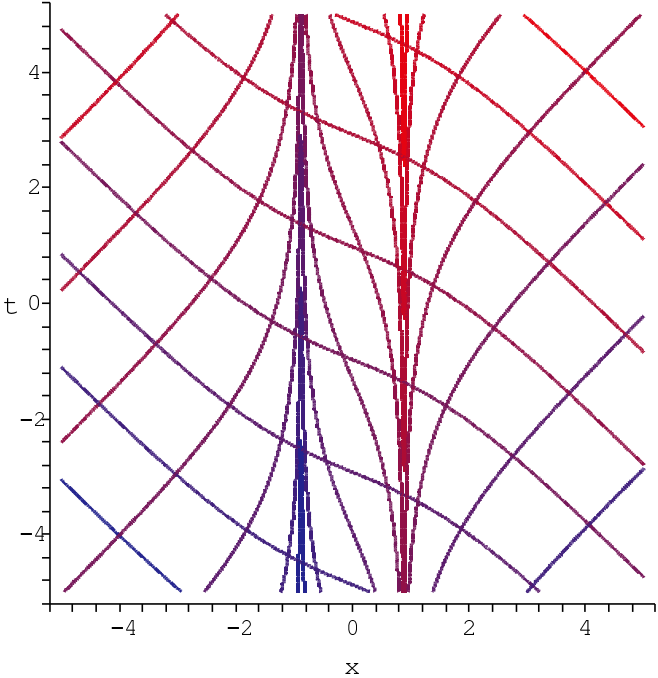}}}
\hfil
}
\bigskip
\caption{
Black hole--white hole configuration. $u=\mathrm{const}$ are the
blue lines, $w=\mathrm{const}$ are the red lines; $\alpha=\sqrt{2}$.%
}
\label{F:UW-bh-wh}%
\end{figure}

%
The null coordinates for this example are:%
\begin{equation}%
u=t-\frac{x}{c}-\frac{\alpha\,a}{c\sqrt{\alpha^2
-1}}\ln\left|\frac{1-\left(\alpha+\sqrt{\alpha^2
-1}\right)\exp\left(-x/a\right)}{1-\left(\alpha
-\sqrt{\alpha^2-1}\right)\exp\left(-x/a\right)}\right|\;;%
\label{U4}%
\end{equation}%
\begin{equation}%
w=t+\frac{x}{c}+\frac{\alpha\,a}{c\sqrt{\alpha^2
-1}}\ln\left|\frac{1+\left(\alpha+\sqrt{\alpha^2
-1}\right)\exp\left(-x/a\right)}{1+\left(\alpha
-\sqrt{\alpha^2-1}\right)\exp\left(-x/a\right)}\right|\;.%
\label{W4}%
\end{equation}%
The acoustic spacetime is shown in figure~\ref{F:UW-bh-wh}.
Of course, from the point of view of observers located in the
asymptotic region on the right ($x\to +\infty$) there is just a
black hole.  However, observers lying in the middle region with
$x_1<x<x_2$ will experience both a future and a past event
horizon, with the corresponding black and white hole regions.%

This is the non-extremal version of the configuration
considered in section~\ref{staticextremal}, which corresponds
to the choice $\alpha=1$. The surface gravity%
\begin{equation}%
\kappa = {c\over a} \; {\sqrt{\alpha^2-1}\over\alpha}%
\end{equation}%
is now different from zero.  This class of acoustic geometries,
and the ring geometry to be considered below, is of particular
interest for any realistic attempt at building acoustic analogue
black holes in the laboratory.%

%
\subsubsection{Ring configuration}%

\begin{figure}[htbp]
\vbox{
\hfil
\scalebox{0.600}{{\includegraphics{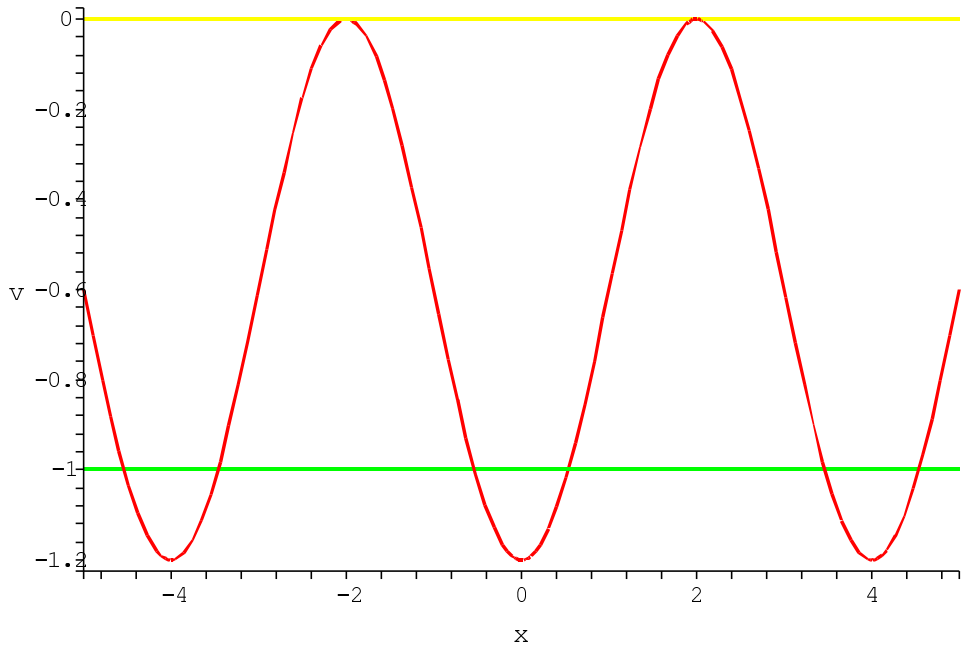}}}
\hfil
}
\bigskip
\caption{
Ring configuration: Velocity profile for a left-going flow,
in a ring with one supersonic region; $\alpha=1.2$; $L=2$.%
}
\label{F:v-ring-bh}
\end{figure}

Consider a left-going flow in a ring.  Assume one supersonic
region, one subsonic, and $v$ bounded. For example, we could take%
\begin{equation}%
v(x) = - \frac{\alpha c}{2}\;\left (1 + \cos\frac{\pi x}{L}\right)%
\label{vring}%
\end{equation}%
with $\alpha>1$ (otherwise there are no sonic points) and $L$
equal to half of the ring length (the ring goes from $-L$ to $L$);
see figure~\ref{F:v-ring-bh}.  This configuration, because
the working fluid is constantly recycled, is perhaps the premier
class of acoustic models that are potentially of experimental
interest~\cite{abh,garay,garay2,laval}.%

%
\begin{figure}[htbp]
\vbox{
\hfil
\scalebox{0.600}{{\includegraphics{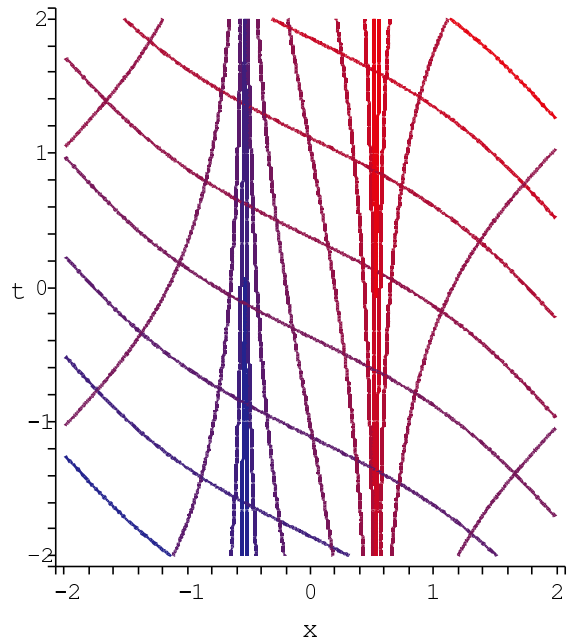}}}
\hfil
}
\bigskip
\caption{
Ring-hole:  Black hole--white hole configuration with
periodic boundary conditions. $u=\mathrm{const}$ are the blue
lines, $w=\mathrm{const}$ are the red lines.  $\alpha=1.2$; $L=2$.
While the technical definition of the absolute horizon
does not work in this situation, the apparent horizons are located
symmetrically about the origin.
}
\label{F:UW-ring-bh}
\end{figure}

%
The null coordinates are%
\begin{equation}%
u = t + \frac{L}{c \pi \sqrt{\alpha - 1}}
    \ln\left|\frac{\sqrt{\alpha - 1}
          + \tan\frac{\pi x}{2 L}}{\sqrt{\alpha - 1}
          - \tan\frac{\pi x}{2 L}}\right |\;,%
\label{Uring}%
\end{equation}%
\begin{equation}%
w = t + \frac{2 L}{c \pi \sqrt{\alpha + 1}}\;{\mathrm{arctan}}
    \left(\frac{1}{\sqrt{\alpha + 1}} \tan\frac{\pi x}{2
    L}\right)\;,%
\label{Wring}%
\end{equation}%
and the acoustic spacetime is shown in figure~\ref{F:UW-ring-bh}.
The black ($+$) and white ($-$) hole horizons occur at%
\begin{equation}%
x_S=\pm\frac{L}{\pi}\,{\mathrm{arccos}}\left(\frac{2}{\alpha}-1\right)%
\end{equation}%
with surface  gravity%
\begin{equation}%
\kappa =  {c \;\pi\;\sqrt{\alpha-1}\over L}\;.%
\end{equation}%
We note that because of the periodicity in space it is not possible to
define absolute horizons (event horizons) in the normal way. Whenever
we speak of horizons in the ring configuration we will be referring
primarily to apparent horizons which we can define in a local manner.
If the apparent horizon asymptotes to a null curve in the infinite
future (or infinite past) then a null curve that is tangent to this
asymptote is the best global definition of a horizon that one could
hope to achieve.

%

\subsubsection{Two black holes}

\begin{figure}[htbp]
\vbox{
\hfil
\scalebox{0.600}{{\includegraphics{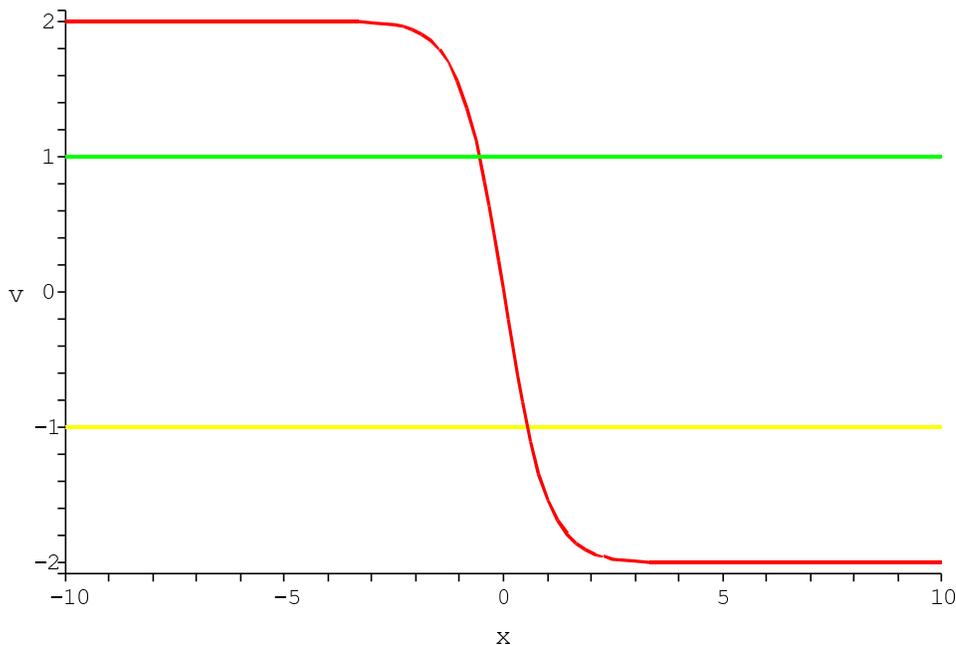}}}
\hfil
}
\bigskip
\caption{
Two black hole configuration: Velocity profile for a left-going
flow, two supersonic regions, one subsonic trapped between them;
$v$ bounded; $\alpha=2$.%
}
\label{F:v-two-wh}
\end{figure}

\begin{figure}[htbp]
\vbox{
\hfil
\scalebox{0.600}{{\includegraphics{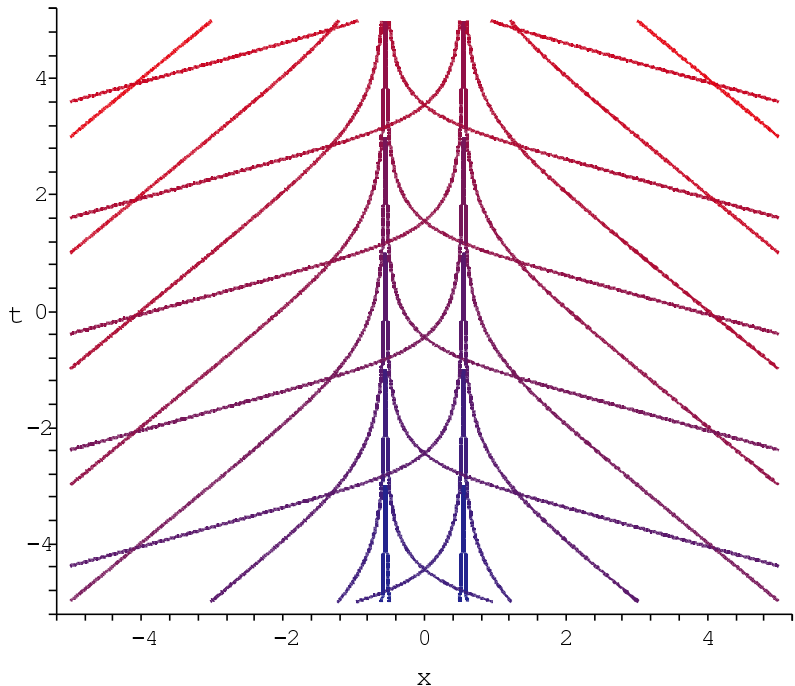}}}
\hfil
}
\bigskip
\caption{
Two black hole configuration. $u=\mathrm{const}$ are the blue
lines, $w=\mathrm{const}$ are the red lines; $\alpha=2$.%
}
\label{F:UW-two-bh}
\end{figure}

As a final example in our collection of stationary examples
we consider a system containing two supersonic regions,
with one subsonic region trapped between them, $v$ remaining
bounded. A simple example of this configuration is given by
the velocity profile%
\begin{equation}%
v(x)=-c\,\alpha\,\tanh\left(x/a\right)\;,%
\label{v6}%
\end{equation}%
with $\alpha>1$; see figure~\ref{F:v-two-wh}.  (To build such a
flow in a condensed matter system one would clearly need to
somehow remove fluid from the vicinity of the origin in order to
be compatible with the equation of continuity. An example of such
behaviour occurs in certain Bose-Einstein condensate-inspired
models of analog black holes where an outcoupling laser beam
focused on the origin is used to systematically destroy the
condensate and set up opposing supersonic flows into the
origin~\cite{abh,garay,garay2}.)%
%
%
%
The null coordinates are%
\begin{equation}%
u=t+\frac{x}{c\left(\alpha
-1\right)}+\frac{\alpha\,a}{c\left(\alpha^2
-1\right)}\ln\left|1-\frac{\alpha+1}{\alpha
-1}\,\exp\left(-2\,x/a\right)\right|\;,%
\label{U6}%
\end{equation}%
\begin{equation}%
w=t+\frac{x}{c\left(\alpha
+1\right)}+\frac{\alpha\,a}{c\left(\alpha^2
-1\right)}\ln\left|1-\frac{\alpha-1}{\alpha
+1}\,\exp\left(-2\,x/a\right)\right|\;,%
\label{W6}%
\end{equation}%
and the spacetime is represented in figure~\ref{F:UW-two-bh}.
The horizons are located at%
\begin{equation}%
x_S=\pm a\ln\left(\frac{\alpha+1}{\alpha-1}\right)%
\end{equation}%
and the surface gravity is%
\begin{equation}%
\kappa = {c\over a}\;\left(\alpha - {1\over\alpha}\right)\;.%
\end{equation}%
%

%

This completes our exhibition of stationary configurations. Additional
examples can be constructed via parity reflection or time reversal but
there are no significant new insights to be gained from such an
exercise.  We shall now turn to the question of classifying
time-dependent acoustic geometries and begin by making some general
comments.


\section{Dynamical zoo}%
\label{sec:coll-zoo}%
\setcounter{equation}{0}%

Let $\sigma(t)$ be some smooth function that increases
monotonically from $0$ (for $t\to -\infty$) to $1$ (for $t\to
+\infty$); for example,%
\begin{equation}%
\sigma(t) = {1\over2}\left[ 1+ \tanh(t/t_0) \right]\;,%
\label{sigma}%
\end{equation}%
where $t_0$ is a positive constant.  Then, for any
$\bar{v}(x)$, the function $v(t,x) := \sigma(t) \,\bar{v}(x)$
represents a velocity profile that changes in time from $0$ to
$\bar{v}(x)$. In particular, we can choose for $\bar{v}(x)$ any
one of the functions considered in the previous examples.%

The differential equations for the worldlines of right-going and
left-going sound rays are, respectively:%
\begin{equation}%
\left(\frac{\d t}{\d x}\right)_R=\frac{1}{c+v(t,x)}\;;%
\label{R}%
\end{equation}%
\begin{equation}%
\left(\frac{\d t}{\d x}\right)_L=-\frac{1}{c-v(t,x)}\;.%
\label{L}%
\end{equation}%
Integrating these equations is not an easy task, even for very
specific and simple choices of the functions $\sigma$ and $\bar{v}$.
However, we can readily identify the apparent acoustic horizons,
where%
\begin{equation}%
\sigma(t)\, \bar{v}(x)=\pm c\;.%
\label{app-hor}%
\end{equation}%
At the apparent horizons, either $\left(\d t/\d x\right)_R$ or
$\left(\d t/\d x\right)_L$ becomes infinite, that is, the signal
world lines (the $u=\mathrm{const}$ and $w=\mathrm{const}$ curves)
have a vertical tangent in the $(t,x)$ diagram.  Other features of
relevance are that when $t\to -\infty$, where we have both
$\left(\d t/\d x\right)_R\to 1/c$ and $\left(\d t/\d x\right)_L\to
-1/c$, the acoustic spacetime becomes Minkowskian.  In contrast,
for $t\to +\infty$, where%
\begin{equation}%
\left(\frac{\d t}{\d x}\right)_R\sim\frac{1}{c+\bar{v}(x)}\;,%
\label{R+}%
\end{equation}%
\begin{equation}%
\left(\frac{\d t}{\d x}\right)_L\sim -\frac{1}{c-\bar{v}(x)}\;,%
\label{L+}%
\end{equation}%
we see that the structure of spacetime resembles those previously
considered in the stationary cases.  Therefore this class of acoustic
geometries can be viewed as providing simple and concrete examples of
black hole and white hole formation in situations where the underlying
physics is very simple and completely under control.  Let us now pass
to more specific descriptions of the behaviour of these dynamical
acoustic geometries.%


\subsection{Flows developing one sonic point}%

\subsubsection{Black hole}%

\begin{figure}[htbp]
\vbox{
\hfil
\scalebox{0.600}{{\includegraphics{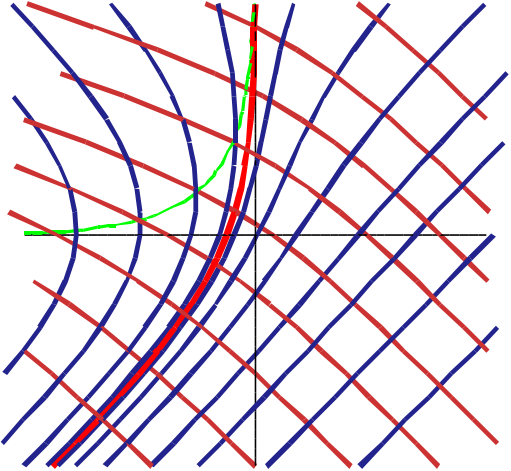}}}
\hfil
}
\bigskip
\caption{
Switching on an acoustic black hole. $u=\mathrm{const}$ are the
blue lines, $w=\mathrm{const}$ are the red lines. The green line
is the apparent horizon. The thick red line is the event horizon.%
}
\label{F:bh-dyn}
\end{figure}

Figure \ref{F:bh-dyn} represents the effect of switching on a
single horizon black hole. As such it is a simple model for what
would in standard general relativity correspond to a collapse
process, and one can easily formulate and analyze questions
regarding the apparent and absolute horizons
and easily verify that they are markedly different.  An
interesting feature is the fact that the absolute horizon 
(the event horizon) extends all the way back to $t=-\infty$ in a
region where the acoustic geometry is approximately flat
(Minkowskian). This is a concrete physical example of the fact
that the absolute horizon, unlike the apparent horizon, is
``teleological''~\cite{hawking,MTW,hell,paradigm}. It is
easy to fall into the trap of thinking that the absolute horizon
``knows'' that supersonic flow will be established sometime in the
future. Instead what is happening is more subtle --- the technical
definition of the absolute horizon requires complete information
of the entire future history of the spacetime, and so the fact
that the absolute horizon extends into regions where there is no
fluid flow is merely a retrodiction, from the point of view
of the infinite future, as to which regions of the spacetime
could have communicated with which parts of future null infinity.
The absolute horizon is not akin to a physical membrane, and is
completely undetectable by any local experiment.  The apparent
horizon, in contrast, can be formulated and detected in
terms of local physics.%

Such behaviour also occurs in standard general relativity, where
it is a reasonably easy exercise to construct spacetimes with
event horizons that first form in regions of the spacetime that
are Riemann-flat, and hence completely curvature-free.%
%

\subsubsection{White hole}%

\begin{figure}[htbp]
\vbox{
\hfil
\scalebox{0.600}{{\includegraphics{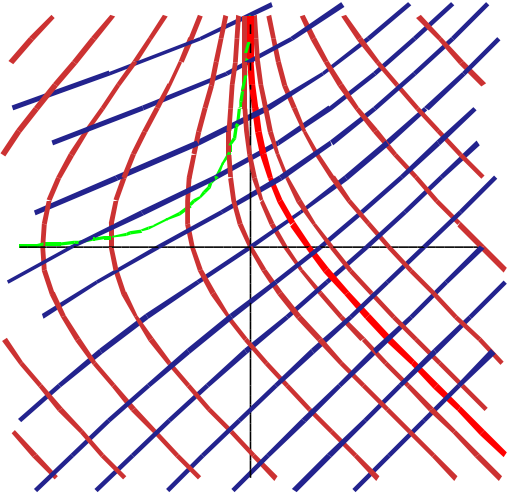}}}
\hfil
}
\bigskip
\caption{
Formation of an acoustic white hole. $u=\mathrm{const}$ are the
blue lines, $w=\mathrm{const}$ are the red lines.  The green line
is the apparent horizon. The thick red line is the event horizon.%
}
\label{F:wh-dyn}
\end{figure}

Similarly, figure~\ref{F:wh-dyn} represents the effect of
switching on a single white hole horizon.  Note that the location
of the apparent horizon is the same as in the black hole case,
while the location of the event horizon is mirror reversed.%

\subsubsection{Black hole, non-physical}%

\begin{figure}[htbp]
\vbox{
\hfil
\scalebox{0.600}{{\includegraphics{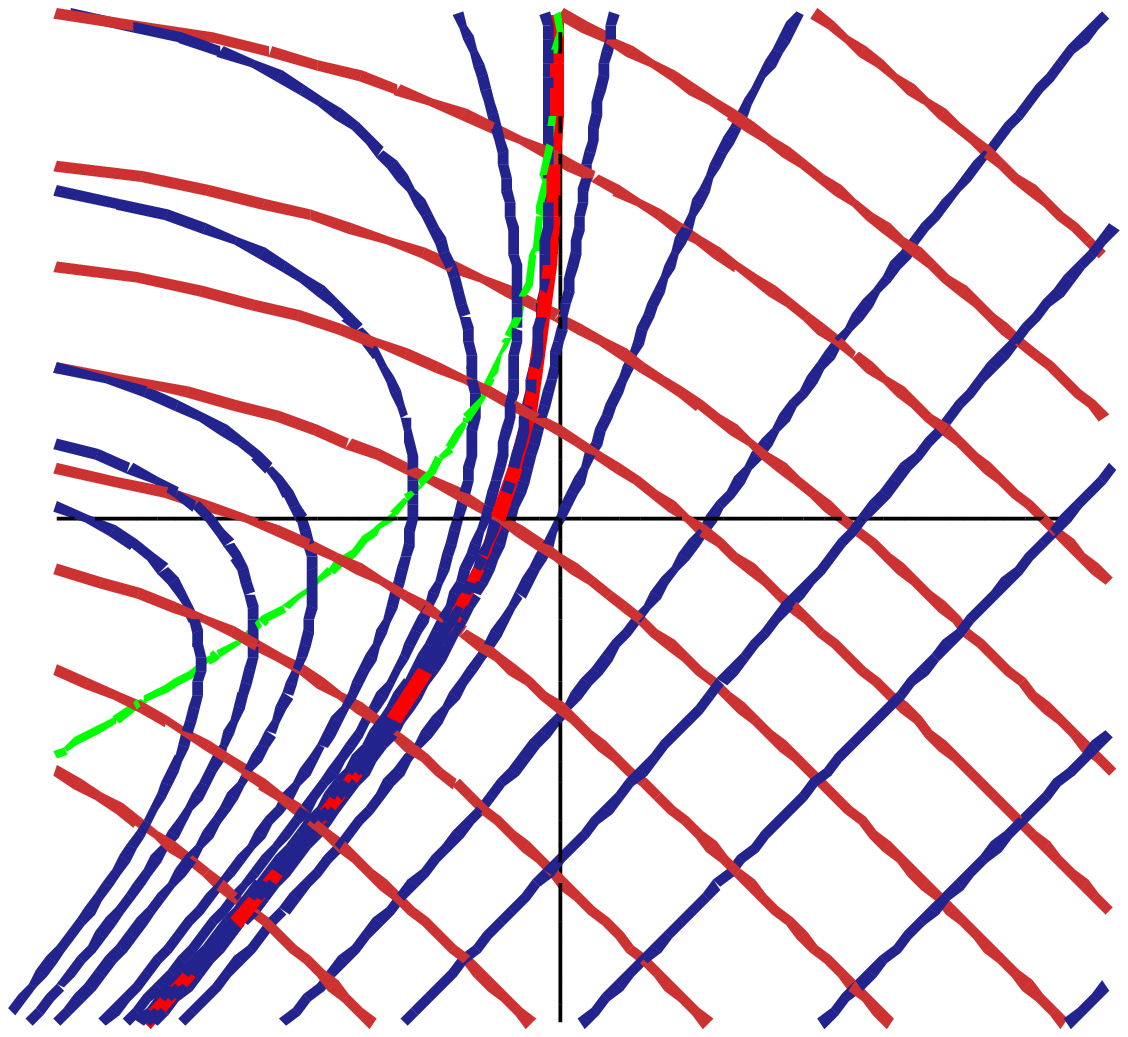}}}
\hfil
}
\bigskip
\caption{
Switching on an unphysical acoustic black hole. $u=\mathrm{const}$
are the blue lines, $w=\mathrm{const}$ are the red lines.  The
green line is the apparent horizon. The thick red line is the
event horizon.%
}
\label{F:unph-bh-dyn}
\end{figure}

The situation in figure~\ref{F:unph-bh-dyn} is ``unphysical'' in
the sense that the fluid velocity goes to infinity at one spatial
limit. Though this is bad from the point of view of a real
physical fluid flow, it is actually rather close to the situation
that arises in a collapse to a Schwarzschild black hole, wherein
the Painlev\'e--Gullstrand ``velocity'' does tend to infinity at
the central singularity.%


\subsubsection{Extremal black hole}%

\begin{figure}[htbp]
\vbox{
\hfil
\scalebox{0.600}{{\includegraphics{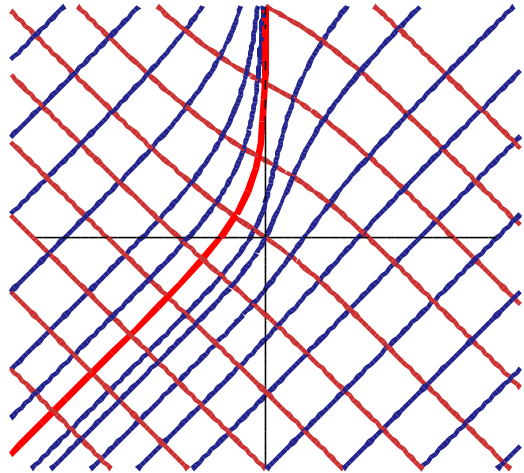}}}
\hfil
}
\bigskip
\caption{
Switching on an extremal black hole.  $u=\mathrm{const}$ are the
blue lines, $w=\mathrm{const}$ are the red lines.  The apparent
horizon is a single isolated point. The thick red line is the event
horizon. This diagram is qualitatively indistinguishable from that
describing the formation of a critical black hole.%
}
\label{F:ex-bh-dyn}
\end{figure}

In figure~\ref{F:ex-bh-dyn} we see the phenomenon of switching on
an extremal black hole. The apparent horizon is now simply an
isolated point in the infinite future, while the location of the
event horizon asymptotes to its known position in the stationary
case.%

%


\subsubsection{Critical black hole}%

Switching on a critical black hole, is qualitatively similar to
switching on an extremal black hole, hence see again
figure~\ref{F:ex-bh-dyn} above.  An interesting point is that in order
for the horizon to form the interpolating function $\sigma(t)$ must
approach unity with sufficient rapidity. In particular the existence
or not of an event horizon on these geometries will depend on the
possibility of sending signals from the left-hand-side that then reach
the right-hand-side infinity.

In order to see this let us start by noticing that the velocity of such
right-moving signals is given by equation~(\ref{R}) and takes the form
\begin{eqnarray}%
{dx \over dt}=c-v(x,t)=c-\sigma(t)\bar v(x).
\end{eqnarray}%
where here $\bar{v}$ is the absolute value of the velocity.  At very
late times $t \to +\infty$ we can write $\sigma(t)=1-A(t)$ where now
$A(t)$ encodes the way $\sigma(t)$ approaches unity. We can then write
\begin{eqnarray}%
{dx \over dt}=c-v(x,t)=A(t)\bar v(x) + c-\bar v(x),
\end{eqnarray}%
but we know that $c-\bar v(x) \geq 0 $ at all times so that 
\begin{eqnarray}%
{dx \over dt} \geq A(t)\bar v(x).
\end{eqnarray}%
Integrating this inequality we have   
\begin{eqnarray}%
\int_{x_i}^{x_\infty} {dx\over \bar v(x)} \geq
\int_{t_i}^{+\infty} A(t)dt,
\end{eqnarray}%
where $x_\infty$ is the location of the right-moving sound ray at time
$t=+\infty$.  Now the convergence or otherwise of the right-hand-side
integral depends on the specific form of $A(t)$. For example, for
$A(t)= N e^{-t/t_0}$ [which is the late-time behaviour we deduce from
the $\sigma(t)$ we chose in (\ref{sigma})] or for another example
taking $A(t)= N/ t^{n}$, with $n > 1$, we have that this
left-hand-side integral yields a finite result.  The inequality then
permits $x_\infty$ to be finite, and therefore we deduce that in these
situations there can be right-moving signals not able to reach the
asymptotic right infinity: An event horizon can be formed (though it
is not guaranteed to be formed by the current argument).  In contrast,
for $A(t)= N t^{-n}$, with $0 < n \leq 1$, this integral diverges and
thus from the inequality we deduce $x_\infty\to+\infty$. So no event
horizon can possibly be formed in the dynamical process. Again, this
might have important implications when analyzing the Hawking process
on these backgrounds.

\subsection{Flows developing two sonic points}%

When we turn to dynamical situations with two sonic horizons we again
note that the behaviour of apparent and absolute horizons can differ
markedly from each other.%

\subsubsection{Black and white hole}%

\begin{figure}[htbp]
\vbox{
\hfil
\scalebox{0.600}{{\includegraphics{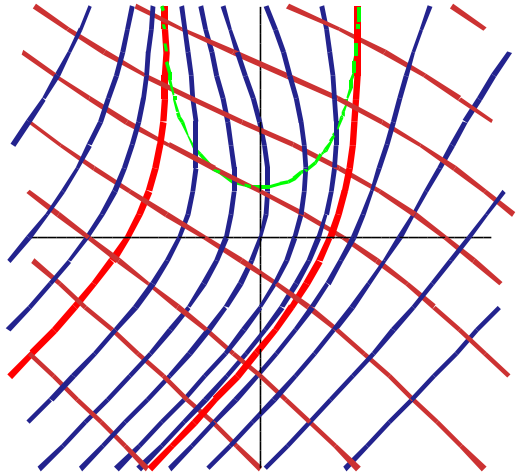}}}
\hfil
}
\bigskip
\caption{
Formation of a black hole--white hole configuration.
$u=\mathrm{const}$ are the blue lines, $w=\mathrm{const}$ are the
red lines.  The green line is the apparent horizon. The two thick
red lines are the event horizons.   $\alpha=\sqrt{2}$.%
}
\label{F:bh-wh-dyn}%
\end{figure}

When switching on a black hole--white hole combination, as
in figure~\ref{F:bh-wh-dyn}, note that the apparent horizon loops
back from one absolute horizon to the other, asymptoting to both
absolute horizons in the far future.%

\subsubsection{Ring configuration}
\begin{figure}[htbp]
\vbox{
\hfil
\scalebox{0.600}{{\includegraphics{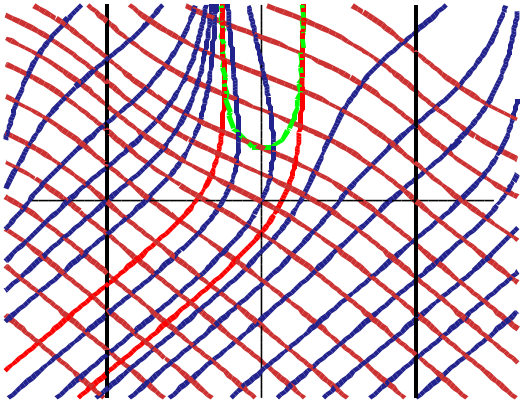}}}
\hfil
}
\bigskip
\caption{
  Formation of a supersonic region in a ring flow.  $u=\mathrm{const}$
  are the blue lines, $w=\mathrm{const}$ are the red lines. The
  apparent horizon is the green line. The two thick red lines are the
  ``event horizons'', defined as null curves that are tangent to the
  asymptotic future limit of the apparent horizons.  $\alpha=1.2$;
  $L=2$.
}
\label{F:ring-bh-dyn}
\end{figure}

\begin{figure}[htbp]
\vbox{
\hfil
\scalebox{0.600}{{\includegraphics{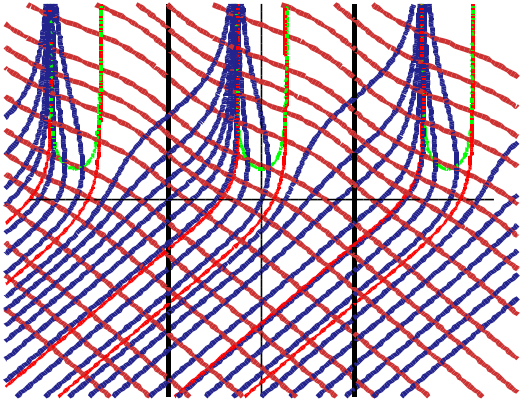}}}
\hfil
}
\bigskip
\caption{
Formation of a supersonic region in a ring flow with
periodicity exposed. $u=\mathrm{const}$ are the blue lines,
$w=\mathrm{const}$ are the red lines. The apparent horizon is the
green line. The thick red lines are the event horizons.
Here, the ``event horizons'' are defined as null curves that are
tangent to the asymptotic future limit of the apparent horizons.
$\alpha=1.2$; $L=2$. 
}
\label{F:ring-bh-dyn-multi}
\end{figure}

In figures~\ref{F:ring-bh-dyn} and \ref{F:ring-bh-dyn-multi} we
see the effect of switching on supersonic flow in a ring geometry.
The first figure emphasises the structure of the fundamental
region, whereas the second diagram emphasises the periodic nature
of the ring geometry.  %

\subsubsection{Two black holes}%

\begin{figure}[htbp]
\vbox{
\hfil
\scalebox{0.600}{{\includegraphics{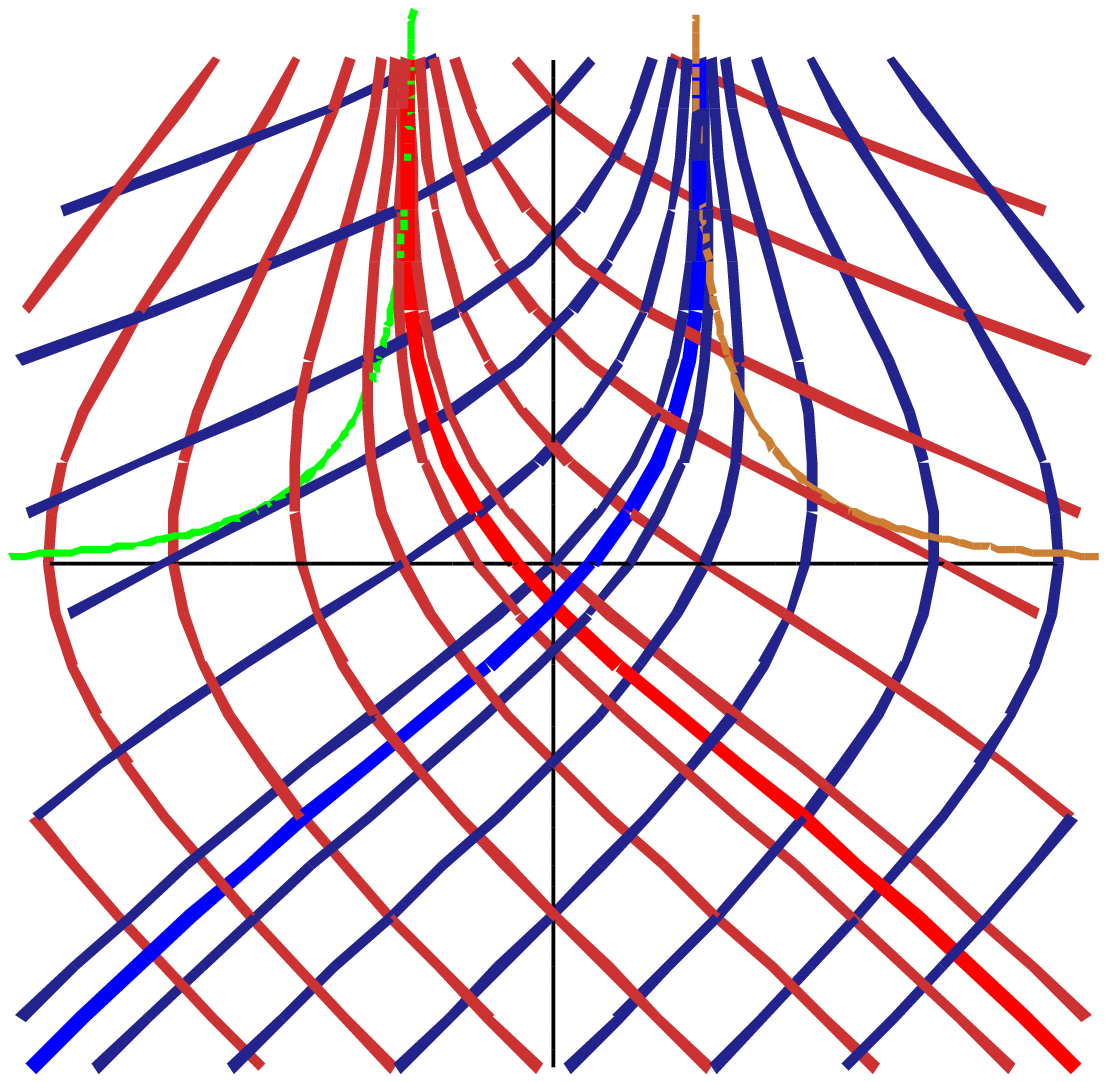}}} 
\hfil
}
\bigskip
\caption{
Formation of a two black hole configuration. $u=\mathrm{const}$
are the blue lines, $w=\mathrm{const}$ are the red lines.  The
green and gold lines are the apparent horizons. The thick red line
and thick blue line are  the two event horizons, which cross at
a finite time. $\alpha=2$.%
}
\label{F:two-bh}
\end{figure}

Note that in figure~\ref{F:two-bh}, where a pair of black holes is
switched on by effectively sucking fluid into the
origin~\cite{abh,garay,garay2},  the geometry is reflection
invariant.%

\section{Compactification --- Stationary geometries}%
\label{sec:compact-stat}%
\setcounter{equation}{0}%

We now present the Penrose--Carter conformal
diagrams~\cite{MTW,hell,penrose,wald} for the various cases in the
previous zoos. These diagrams are used to exhibit the global
causal structure of the model spacetimes we are considering. The
basic idea underneath the conformal diagram of any non-compact 1+1
manifold is that its metric can always be conformally mapped
to the metric of a compact geometry, with a boundary added to
represent events at infinity. Since compact spacetimes are in some
sense ``finite'', they can then properly be drawn on a sheet
of paper.%

To accomplish this, one way is to use coordinates that are 
regular across the horizons and that reach finite values at infinity.
The null coordinates $u$ and $w$ are
clearly unsuitable, since they diverge when $v=-c$ or $v=c$,
respectively, as seen in section~\ref{sec:null}.  However, we can
introduce new coordinates in the same way in which one defines
Kruskal coordinates for Schwarzschild spacetime
\cite{MTW,hell,wald}.  Let $U(u)$ and $W(w)$ be new null
coordinates. The metric (\ref{metric-UW}) with $F=G=1$ becomes%
\begin{equation}%
\g=-\Omega^2\,\left(c^2-v^2\right)\,\frac{\d u}{\d U}\,
\frac{\d w}{\d W}\,\d U\d W\;,%
\label{metric-uw}%
\end{equation}%
where everything is expressed in terms of $U$ and $W$ as
independent variables.  Then, the strategy is to choose the
functions $U(u)$ and $W(w)$ in such a way that the coefficient%
\begin{equation}%
\Omega^2\,\left(c^2-v^2\right)\,\frac{\d u}{\d U}\,\frac{\d w}{\d W}%
\label{guw}%
\end{equation}%
be regular everywhere.%

For example, consider a flow which has a single sonic point
with $v=-c$ at $x=0$, like the one  of
section~\ref{subsubsec:bh}. From equations (\ref{U-asympt}) we find%
\begin{equation}%
w-u\sim\frac{x}{2\,c}+\frac{1}{\kappa}\,\ln|x|\;.%
\label{W-U}%
\end{equation}%
As we approach $x=0$ the coordinate $u$ grows indefinitely, while
$w$ remains finite.  Therefore $w-u\sim -u$ and equation
(\ref{W-U}) can be simplified near the sonic point as%
\begin{equation}%
|x|\sim\exp\left(-\kappa\,u\right)\;.%
\label{X(U)}%
\end{equation}%
Then, near $x=0$ we have%
\begin{equation}%
c^2-v^2\sim 2\,\epsilon\,c\,\kappa\,x\sim
2\,\epsilon\,c\,\kappa\,\frac{x}{|x|}\,\exp\left(-\kappa\,u\right)\;,
\label{c2-v2}%
\end{equation}%
where equation (\ref{v}) has been used in the first step.
Hence, in order to keep the coefficient (\ref{guw}) finite, it is
sufficient that $\d U/\d u$ be proportional to the
expression on the right-hand side of equation (\ref{c2-v2}).
Thus we choose, near the sonic point $x=0$,%
\begin{equation}%
U(u)\propto -\frac{x}{|x|}\,\exp\left(-\kappa\,u\right)\;,%
\label{u}%
\end{equation}%
with a positive proportionality factor, so that $U$ varies
regularly from $-\infty$ to $+\infty$ as $x$ varies from $+\infty$
to $-\infty$.  On the other hand, $w$ is regular everywhere, so we
can choose for $W$ any regular monotonic function of $w$ such that
$W(\pm\infty)=\pm\infty$.%

Let us now choose the new coordinates%
\begin{equation}%
{\cal U}:=\arctan U, ~~~~ {\cal W}:=\arctan W,%
\end{equation}%
that take values in the range $(-\pi/2,\pi/2)$.  The acoustic metric
is
\begin{equation}%
\g= - \Omega^2 \; \frac{c^2-v^2}{\cos^2{\cal U}\,\cos^2{\cal W}}\,\frac{\d
u}{\d U}\,\frac{\d w}{\d W}\,\d {\cal U}\,\d {\cal W}\;,%
\end{equation}%
where everything is expressed as a function of $\cal U$ and $\cal
W$.  Hence, the metric is conformal to%
\begin{equation}%
\bar{\g} = - \d {\cal U}\,\d {\cal W}\;,%
\end{equation}%
so the causal structure of the spacetime is identical to the one
of the portion $({\cal U},{\cal W}) \in (-\pi/2,\pi/2)\times
(-\pi/2,\pi/2)$ of Minkowski spacetime.  The points on the
boundary, where $\cal U$ and $\cal W$ attain values $\pm\pi/2$,
represent sonic boundaries or points at infinity in the 
acoustic spacetime.  We now
consider the Penrose--Carter conformal diagrams for the various
acoustic geometries we have investigated.%

\subsection{Flows with one sonic point}%

\subsubsection{Black hole}%

\begin{figure}[htbp]
\vbox{
\hfil
\scalebox{0.600}{{\includegraphics{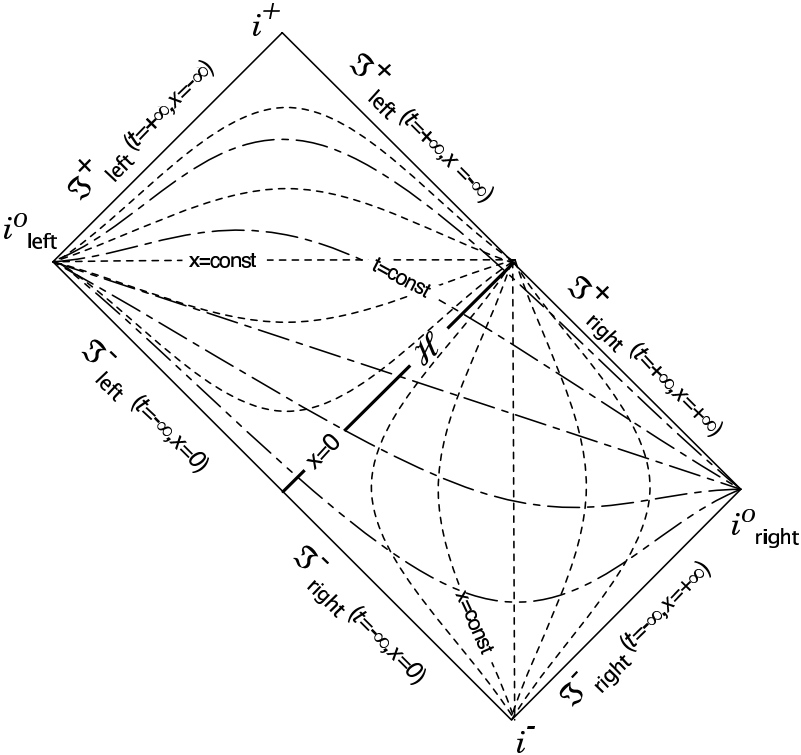}}}
\hfil
}
\bigskip
\caption{
Conformal diagram of an acoustic black hole.%
}
\label{F:conf-bh}
\end{figure}

For the case of a single isolated black hole horizon we find the
Penrose--Carter diagram of figure~\ref{F:conf-bh}.\footnote{In the
figure we have introduced an aspect ratio different from unity for the
coordinates $\cal U$ and $\cal W$, in order to make the various
regions of interest graphically more clear.} As we have already
commented, in the acoustic spacetimes with no periodic identifications
there are two clearly differentiated notions of asymptotia, ``right''
and ``left''. In all our figures we have used subscripts ``right'' and
``left'' to label the different null and spacelike infinities.
In addition, we have denoted the different sonic-point boundaries
with ${\Im}^{\pm}_\mathrm{right}$ or ${\Im}^{\pm}_\mathrm{left}$
depending on whether they are the starting point ($-$ sign) or the ending
point (+ sign) of the null geodesics in the right or left parts of the 
diagram.
  
In contradistinction to the Penrose--Carter diagram for the
Schwarzschild black hole (which in the current context would have to
be an eternal black hole, not one formed via astrophysical stellar
collapse) there is no singularity. On reflection, this feature of the
conformal diagram should be obvious since the fluid flow underlying
the acoustic geometry is nowhere singular.%
%
%
Note that the event horizon $\cal{H}$ is the boundary of the
causal past of future right null infinity; that is, ${\cal
H} = \dot{J}^-({\Im}^+_\mathrm{right})$, with standard
notations \cite{MTW,hell,wald}.%

\subsubsection{White hole}%

\begin{figure}[htbp]
\vbox{
\hfil
\scalebox{0.600}{{\includegraphics{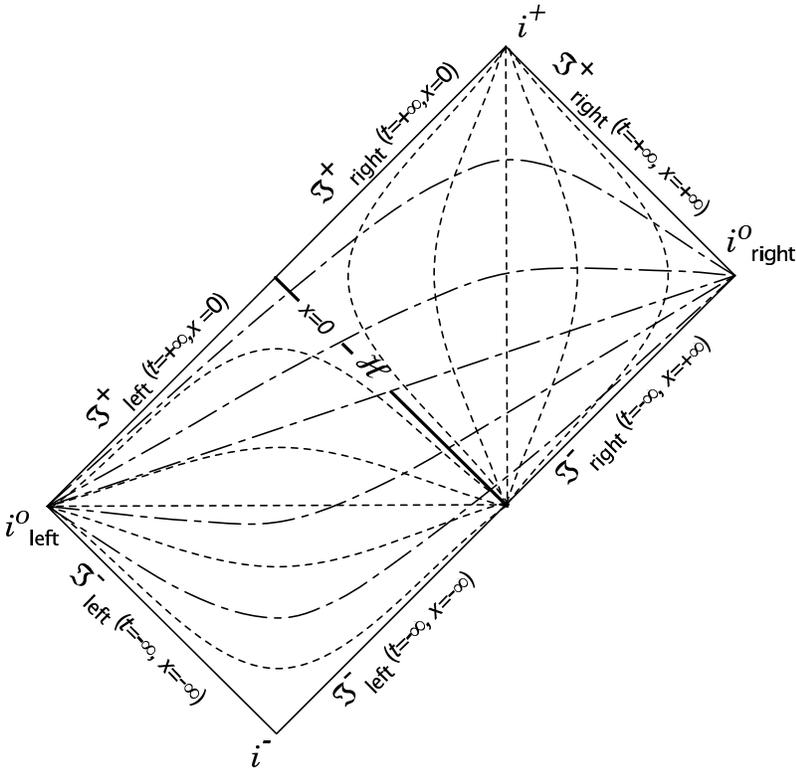}}}
\hfil
}
\bigskip
\caption{
Conformal diagram of an acoustic white hole.%
}
\label{F:conf-wh}
\end{figure}

The Penrose--Carter diagram for a single white hole horizon, as presented
in figure~\ref{F:conf-wh}, does not provide any new surprises.
Again note the absence of any singularity. Though this is a simple
observation we feel it should be emphasised: These last two very
simple examples conclusively demonstrate that the existence of
black hole and white hole absolute horizons (event horizons) does
not require the presence of any geometrical singularity anywhere
in the spacetime.  That event horizons in general relativity are
always related to the presence of some sort of curvature
singularity is a geometrodynamic statement that depends
specifically on the use of the vacuum Einstein equations, or
on the use of specific matter models plus the Einstein
equations~\cite{hell}. Since the geometry of acoustic
spacetimes is not governed by the Einstein equations, we can very
easily come up with global causal structures that are more general
than those encountered within the context of standard
general relativity. The utility of doing so is twofold:%
\begin{itemize}%
\item If one wishes to physically build an acoustic spacetime, say
with a view to experimentally probing curved-space quantum field
theory~\cite{Unruh1,abh,laval,fischer,bec,silke,silke2,frw,frw2},
it is very useful to know how these spacetimes can differ from
those in  standard general relativity.%
\item If in contrast one is interested in studying extensions to
standard general relativity, then these particularly simple
acoustic spacetimes provide concrete examples that can be used as
a guide to the types of generalization that are in principle
possible.%
\end{itemize}%
%
%
Note that in this situation  the event horizon $\cal{H}$ is the
boundary of the causal future of past right null
infinity, so that ${\cal H} = \dot{J}^+({\Im}^-_\mathrm{right})$.%

\subsubsection{Black hole, non-physical}%

\begin{figure}[htbp]
\vbox{
\hfil
\scalebox{0.600}{{\includegraphics{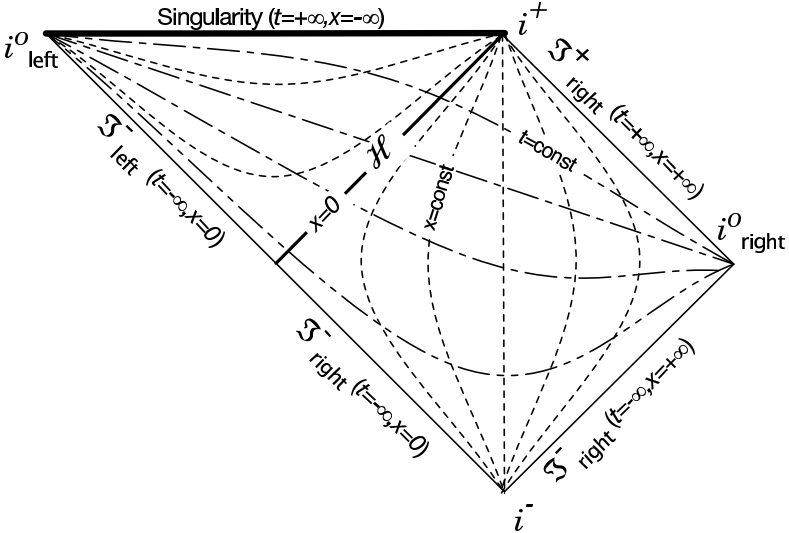}}}
\hfil
}
\bigskip
\caption{
Conformal diagram of an unphysical acoustic black hole.  The
thick horizontal line corresponding to $x=-\infty$ is a
spacelike singularity.%
}
\label{F:conf-unph-bh}
\end{figure}

The ``non-physical'' acoustic black hole, because it does possess
an infinite velocity singularity in the fluid flow, now does have
the potential of containing a curvature singularity in the
spacetime geometry.  Indeed, inspection of the Penrose--Carter diagram in
figure~\ref{F:conf-unph-bh} confirms the presence of a
spacelike singularity.%
%

\subsubsection{Extremal black hole}%

\begin{figure}[htbp]
\vbox{
\hfil
\scalebox{0.600}{{\includegraphics{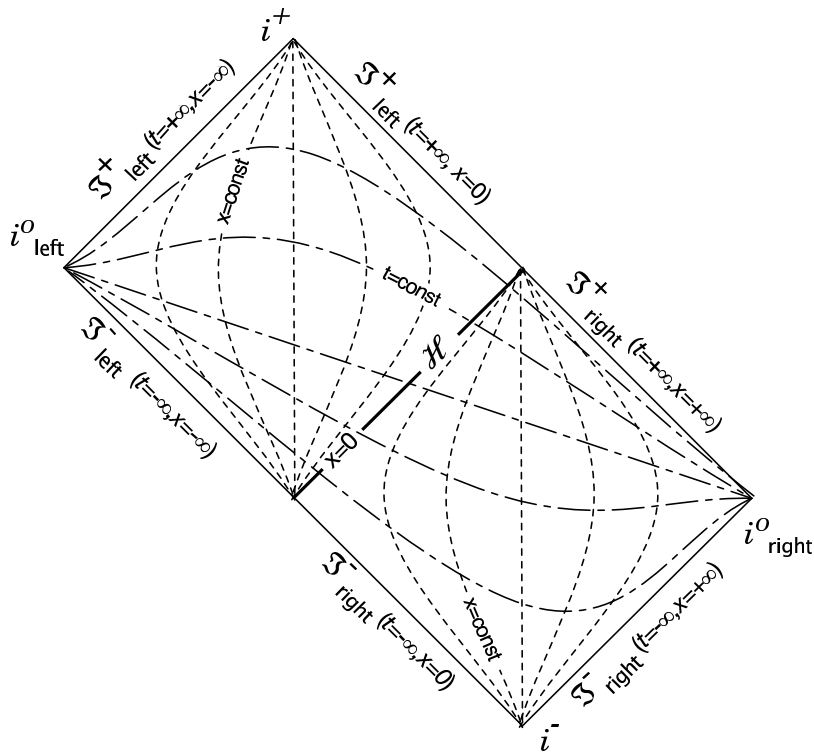}}}
\hfil
}
\bigskip
\caption{
Conformal diagram of an acoustic extremal black hole.  This is
identical to the conformal diagram of an acoustic critical black
hole.%
%
}
\label{F:conf-ex-bh}
\end{figure}

The global causal structure of the extremal acoustic black hole, as
represented in figure~\ref{F:conf-ex-bh}, is rather similar (but not
identical) to that for a generic acoustic black hole. It is also
identical to the causal structure of the critical black hole
considered below. Differently from gravitational black holes, the
major differences between extremal and non-extremal configurations are
in this case geometrical rather than topological. In this sense the
extremal limit for acoustic spacetimes appears to be better behaved
than that for general relativistic spacetimes. Actually the absence of
topological features of extremal spacetimes in acoustic geometries can
be regarded as another evidence of the fact that in acoustic
geometries the conceptual degeneracies of general relativity are often
solved.  For instance, the Penrose--Carter diagram for the extremal
black hole solution of general relativity is topologically different
from that for the non-extremal case and this is indeed the origin of
the possibility to formulate a third law of thermodynamics for black
holes. In this case the absence of Hawking temperature seems tied up
with a non-trivial change in the topological structure of the
spacetime with deep consequences on the black hole dynamics. However
the analog acoustic spacetime is showing us that the the existence (or
otherwise) of Hawking radiation is a purely kinematic feature that is
independent of the dynamics underlying the evolution of the spacetime
(i.e. on the Einstein equations). This is why in these acoustic
geometries the temperature effects are not directly tied up with
entropy considerations as in the four laws of black hole
mechanics~\cite{Unruh1,Visser3,essential}.

\subsubsection{Critical black hole}

The critical acoustic black hole has causal structure identical to
the extremal black hole,  already represented in
figure~\ref{F:conf-ex-bh}.  The differences are again geometrical,
not topological, and do not show up at the level of global causal
structure.%

\subsection{Flows with two sonic points}%

\subsubsection{Black and white hole}%

\begin{figure}[htbp]
\vbox{
\hfil
\scalebox{0.600}{{\includegraphics{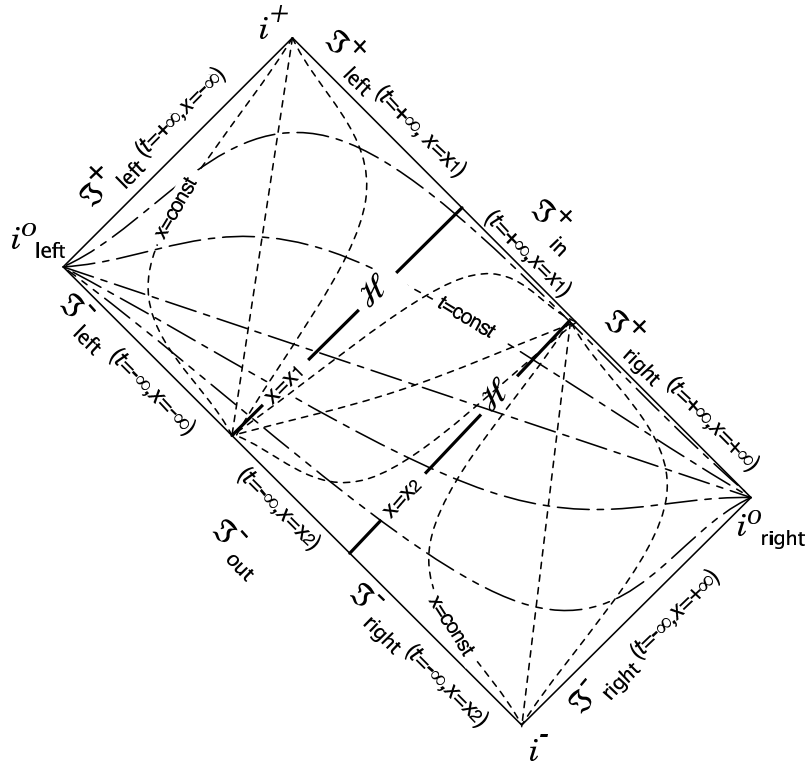}}}
\hfil
}
\bigskip
\caption{
Conformal diagram of an acoustic black hole--white hole pair. Note
the complete absence of singularities. 
Furthermore as $x_2\to x_1$ the two horizons
coalesce, and the causal structure smoothly limits to that of the
extremal black hole in figure~\ref{F:conf-ex-bh}.%
}
\label{F:conf-bh-wh}
\end{figure}

The acoustic geometry for a black hole--white hole
combination again has no singularities in the fluid flow and no
singularities in the spacetime curvature. In particular, from
figure~\ref{F:conf-bh-wh}, we note the complete absence of
singularities. Furthermore as $x_2\to x_1$, so that the two
horizons coalesce, the causal structure smoothly limits to that of
the extremal  black hole described in figure~\ref{F:conf-ex-bh}
above.  This behaviour is in marked contrast to what typically
happens in standard general relativity, where the extremal limit
is quite singular.%

\subsubsection{Ring configuration}%

\begin{figure}[htbp]
\vbox{
\hfil
\scalebox{0.600}{{\includegraphics{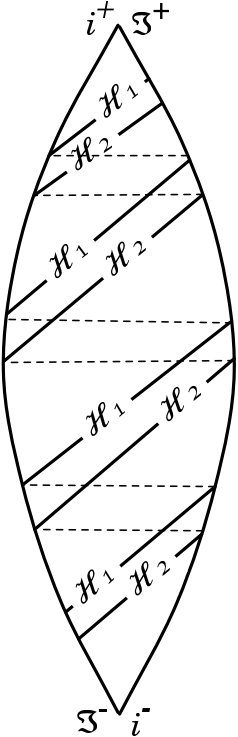}}}
\hfil
}
\bigskip
\caption{
Ring hole: Conformal diagram of an acoustic black hole--white hole
pair in a ring.   
}
\label{F:conf-ring-bh}
\end{figure}

The ring configuration represented in
figure~\ref{F:conf-ring-bh} is effectively a black hole--white
hole combination subject to periodic boundary conditions.  %
As such it cannot strictly speaking be represented on a single
flat piece of paper but must be drawn on a cylindrical sheet of
paper.  For practical purposes however a single flat sheet of
paper with suitable identifications suffices.%

%

%


\subsubsection{Two black holes}%

\begin{figure}[htbp]
\vbox{
\hfil
\scalebox{0.600}{{\includegraphics{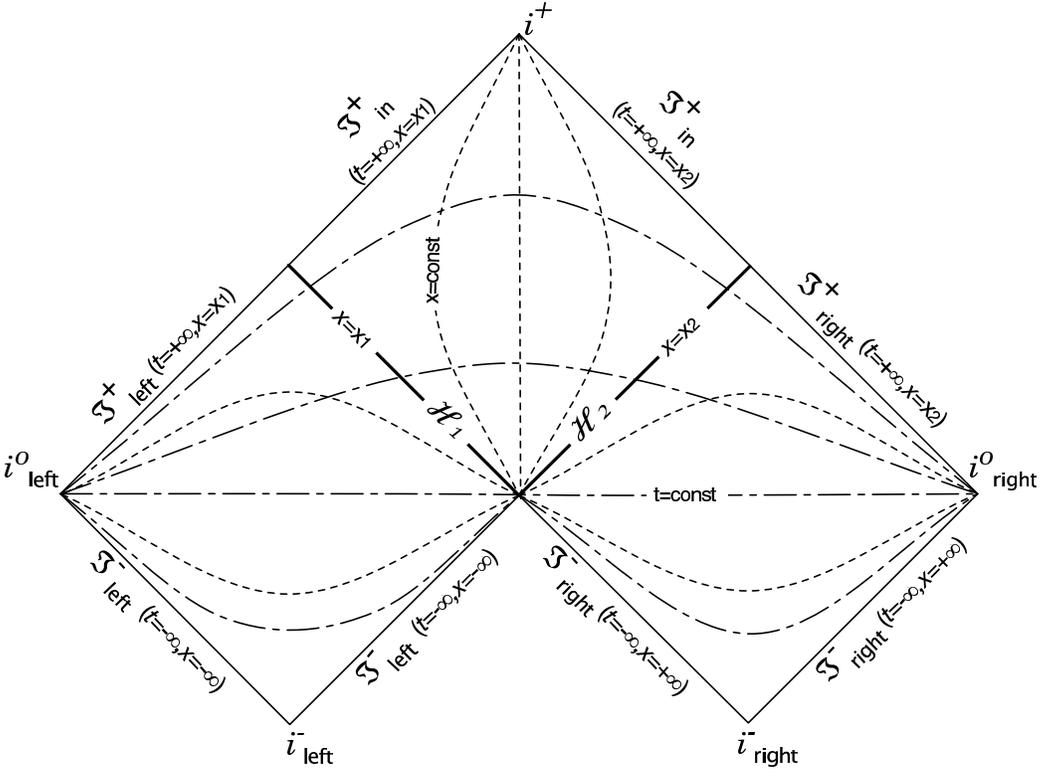}}}
\hfil
}
\bigskip
\caption{
Conformal diagram of a flux with two acoustic black holes.
%
}
\label{F:conf-two-bh}
\end{figure}

For our final example of a stationary acoustic geometry we
consider the case of two black hole horizons.  From
figure~\ref{F:conf-two-bh} we again note the complete absence of
any spacetime singularity, despite the presence of two
well-defined symmetrically placed event horizons.%
%


\section{Compactification --- Dynamical geometries}%
\label{sec:compact-dyn}%
\setcounter{equation}{0}%

Let us now consider the Penrose--Carter diagrams for the global
causal structure of time-dependent acoustic geometries. In all the
cases considered below the geometry is trivial in the infinite
past (flat Minkowski space), but has some nontrivial causal
structure in the infinite future, with the concomitant presence of
one or more event horizons.%

\subsection{Flows developing one sonic point}%

\subsubsection{Black hole}%

\begin{figure}[htbp]
\vbox{
\hfil
\scalebox{0.600}{{\includegraphics{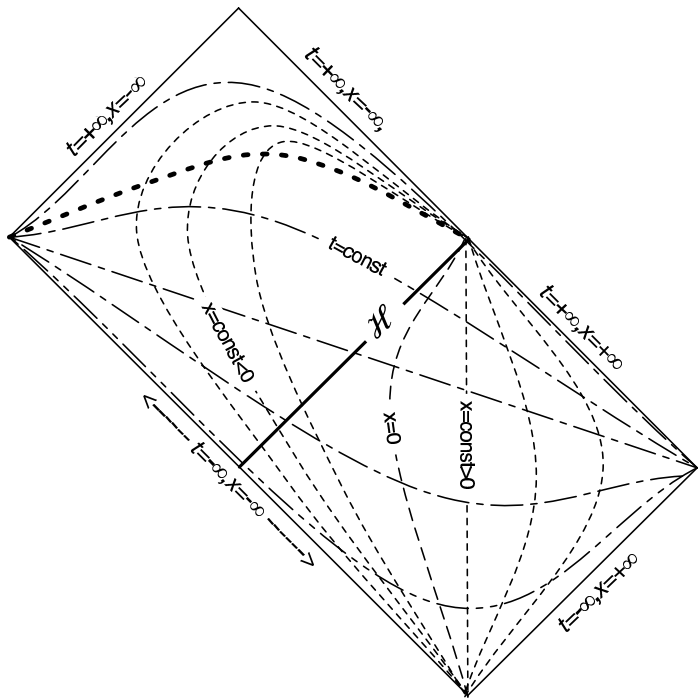}}}
\hfil
}
\bigskip
\caption{
Conformal diagram of the switching on of an acoustic black hole.
Note that the lines $x=\mbox{const}$ become null at the apparent
horizon (dashed line).%
}
\label{F:conf-bh-dyn}
\end{figure}

Formation of an acoustic black hole by switching on a fluid flow
is represented in figure~\ref{F:conf-bh-dyn}.  This is the
analogue, in the acoustic geometries, of an astrophysical black
hole formed by stellar collapse. 
%


\subsubsection{White hole}%

\begin{figure}[htbp]
\vbox{
\hfil
\scalebox{0.600}{{\includegraphics{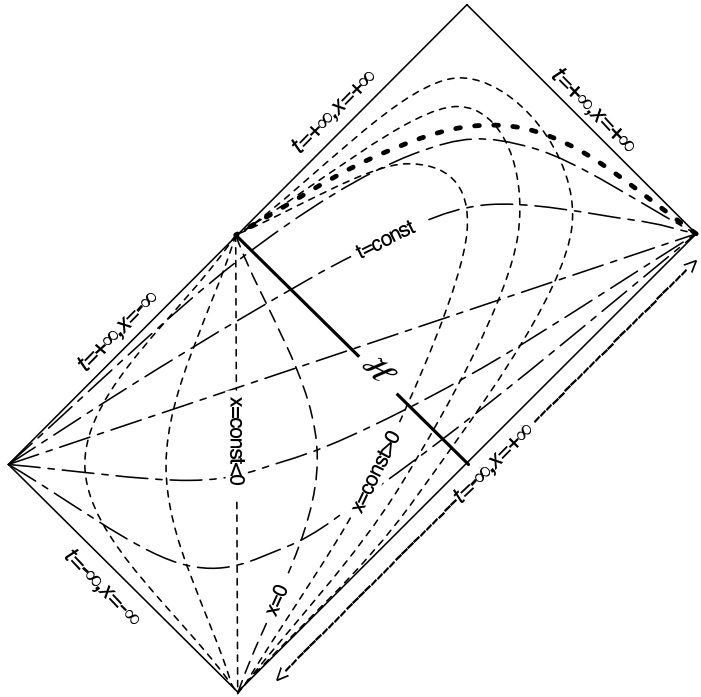}}}
\hfil
}
\bigskip
\caption{
Conformal diagram of the formation of an acoustic white hole.
}
\label{F:conf-wh-dyn}
\end{figure}

In figure~\ref{F:conf-wh-dyn} we present the Penrose--Carter diagram of a
white hole that is switched on as the fluid flow is
accelerated from zero to its final partially supersonic flow.%


\subsubsection{Black hole, non-physical}%

\begin{figure}[htbp]
\vbox{
\hfil
\scalebox{0.600}{{\includegraphics{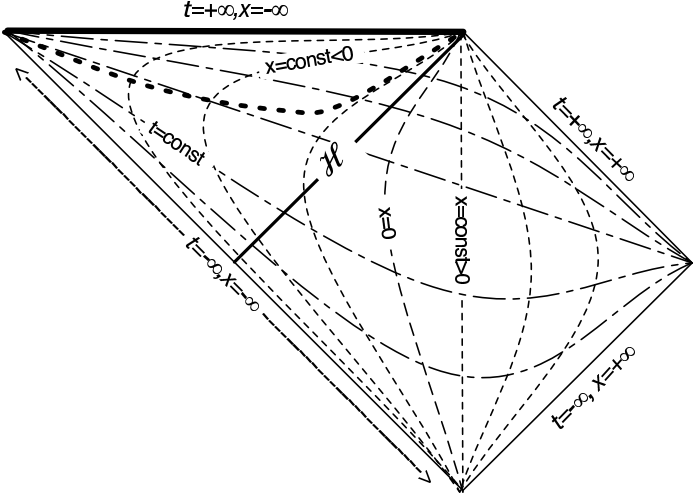}}}
\hfil
}
\bigskip
\caption{
Conformal diagram of the formation of an unphysical acoustic black
hole.%
}
\label{F:conf-unph-bh-dyn}
\end{figure}

For this geometry, represented by figure~\ref{F:conf-unph-bh-dyn},
there is a region of infinite velocity flow at left spatial infinity
($x\to-\infty$) for all finite times. This indicates the existence
of a spacetime singularity that reaches back asymptotically to the
infinite past. %


\subsubsection{Extremal black hole}%

\begin{figure}[htbp]
\vbox{
\hfil
\scalebox{0.600}{{\includegraphics{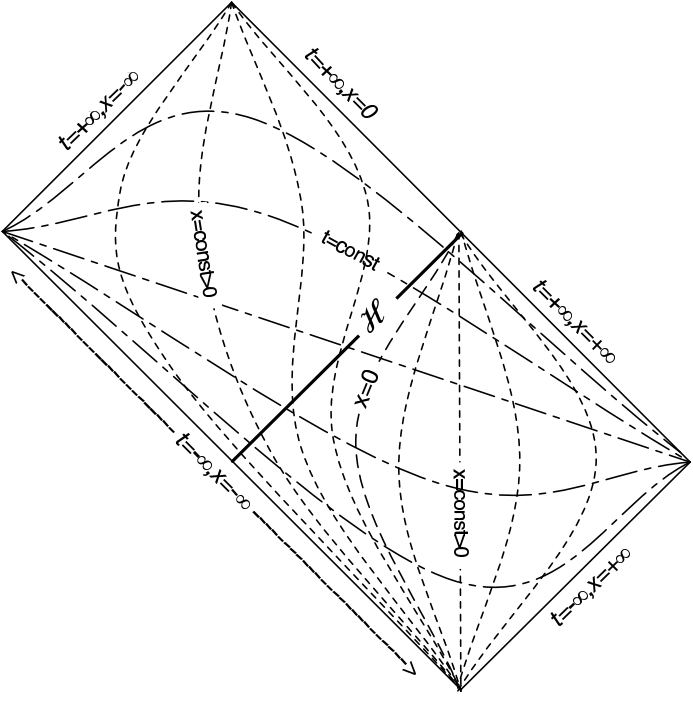}}}
\hfil
}
\bigskip
\caption{
Conformal diagram of the formation of an acoustic extremal black
hole. The causal structure is identical to that for the formation
of an acoustic critical black hole.%
}
\label{F:conf-ex-bh-dyn}
\end{figure}

In figure~\ref{F:conf-ex-bh-dyn} we present the Penrose--Carter diagram
for the causal structure of an extremal black hole that is
switched on as a fluid flow is accelerated from zero velocity to
exactly reach the speed of sound. Note the strong resemblance to
the case of switching on a single black hole, even though in this
case the apparent horizon is a single point at future infinity.%


\subsubsection{Critical black hole}

The qualitative features of the Penrose--Carter diagram for the
formation of a critical black hole are actually identical to that for
formation of an extremal black hole, as presented in
figure~\ref{F:conf-ex-bh-dyn}. The differences lie in the details of
the spacetime geometry, but global causal structures are
identical.%

\subsection{Flows developing two sonic points}

By now the general pattern of the results should be clear. Adding time
dependence so that we can explicitly discuss black hole formation in
these acoustic geometries has complicated the spacetime geometry
but has not led to massive modifications of the qualitative causal
structure.%

\subsubsection{Black and white hole}%

\begin{figure}[htbp]
\vbox{
\hfil
\scalebox{0.600}{{\includegraphics{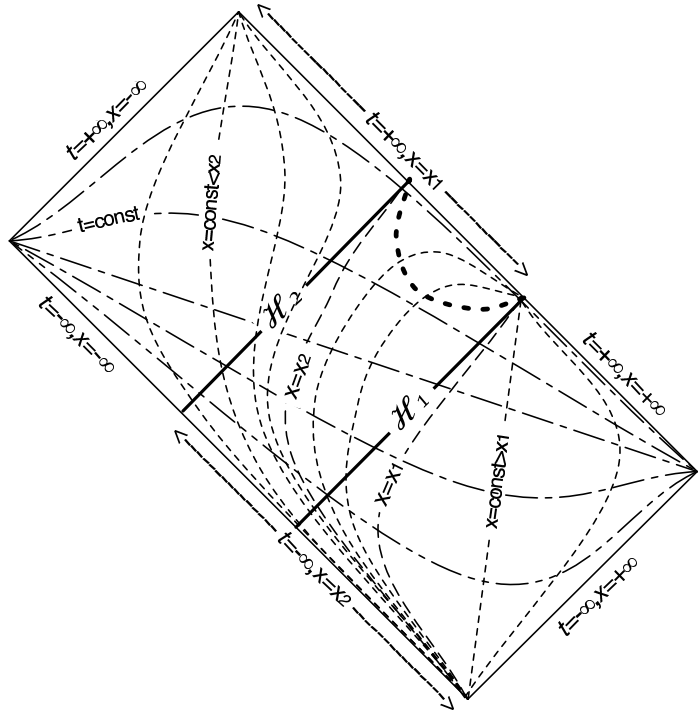}}}
\hfil
}
\bigskip
\caption{
Conformal diagram of the formation of an acoustic black
hole--white hole pair.%
}
\label{F:conf-bh-wh-dyn}
\end{figure}

When a black hole--white hole pair is formed, as in
figure~\ref{F:conf-bh-wh-dyn}, the major change compared to an
eternal black hole--white hole pair lies in the behaviour of
the apparent horizon.%


\subsubsection{Ring geometry}%

\begin{figure}[htbp]
\vbox{
\hfil
\scalebox{0.600}{{\includegraphics{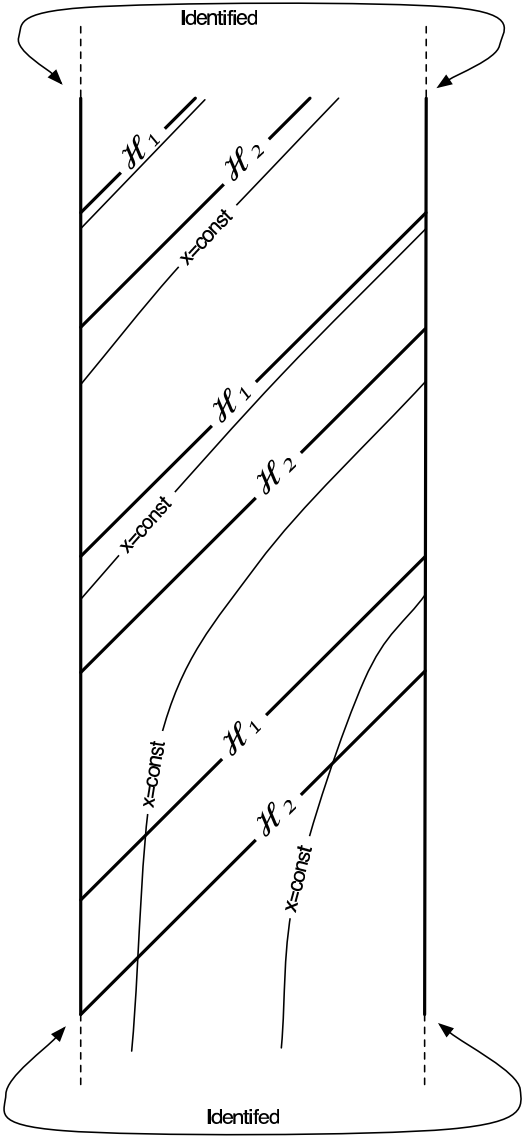}}}
\hfil}
\bigskip
\caption{
Diagram of the formation of an acoustic black hole/white hole
pair in a ring, before compactification.  %
}
\label{F:conf-ring-bh-wh-dyn-2}
\end{figure}

\begin{figure}[htbp]
\vbox{
\hfil
\scalebox{0.600}{{\includegraphics{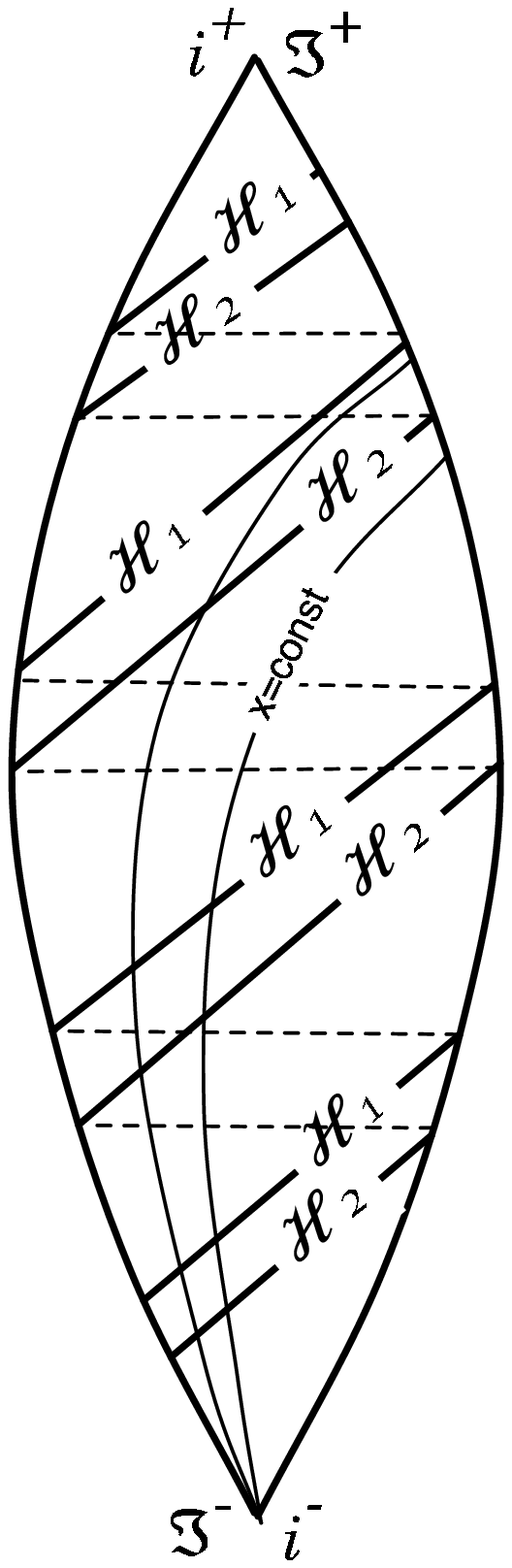}}}
\hfil}
\bigskip
\caption{
Conformal diagram of the formation of an acoustic black hole/white
hole pair in a ring. %
}
\label{F:conf-ring-bh-wh-dyn-3}
\end{figure}

Switching on a black hole/white hole combination in the ring
geometry generates the causal structure presented in
figures~\ref{F:conf-ring-bh-wh-dyn-2} and
\ref{F:conf-ring-bh-wh-dyn-3}.  %



\subsubsection{Two black holes}

\begin{figure}[htbp]
\vbox{
\hfil
\scalebox{0.600}{{\includegraphics{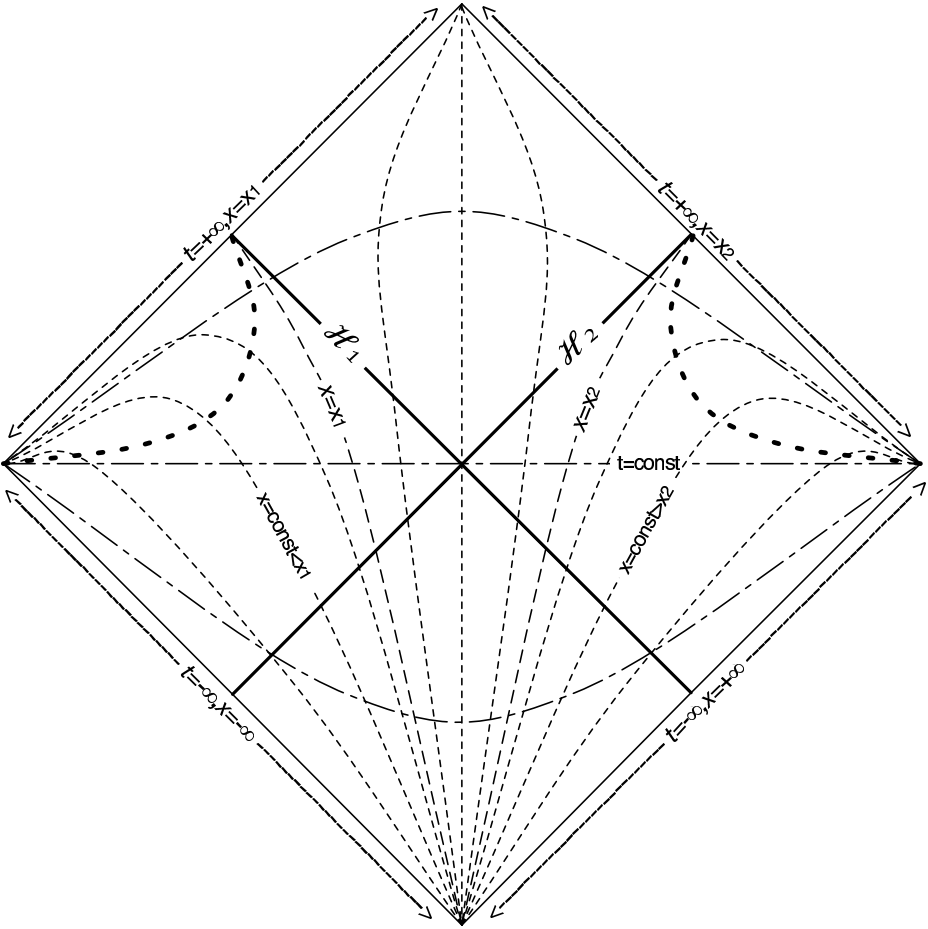}}}
\hfil
}
\bigskip
\caption{
Conformal diagram of the generation of a fluid flow with two acoustic
black holes.  %
}
\label{F:conf-two-bh-dyn}
\end{figure}

The switching on of paired black holes produces the pleasingly
symmetric Penrose--Carter diagram of figure~\ref{F:conf-two-bh-dyn}. Note
the complete absence of any singularities and the presence of a
bifurcation point where the two horizons intersect.  This
spacetime is not time reversal invariant, and serves as an
explicit counterexample to the mistaken idea that a bifurcate
horizon is necessarily Killing.%

\clearpage 

\section{Analytic extension}%
\label{sec:regular}%
\setcounter{equation}{0}%

After describing the causal structures for our catalog of acoustic
spacetimes, we will now pass to discuss the mathematical possibility
of extending these geometries beyond their boundaries. As we commented
in section~\ref{S:fundamental}, when one analyses physical acoustics
in 2+1 and 3+1 dimensions (from which our 1+1 geometries follow by
reduction), one arrives at a specific acoustic metric and not to an
equivalence class of conformally related metrics.  This is the point
of view we adopt in this section. In particular, the following
discussion assumes an acoustic metric of the form (\ref{metric-UW})
with constant $c$.  For example, the acoustic metric that one
naturally finds for a barotropic fluid in 3+1 dimensions is
(\ref{metric-UW}) multiplied by a specific conformal factor
$\Omega^2=\rho(c)/c$, that in the simplest case in which $c=$constant
becomes constant and so irrelevant\footnote{Actually, 
in 1+1 dimensions the continuity equation gives $\rho v=\mbox{const}$
so that $\rho(v)\propto 1/v$ is not constant (although at least it is
finite at the sonic points). In contrast, in 3+1 dimensions with
appropriate symmetry (such as a 1-dimensional duct with variable cross
sections) we certainly can arrange $c=$constant.}.

\begin{figure}[htbp]
\vbox{ \hfil
\scalebox{0.600}{{\includegraphics{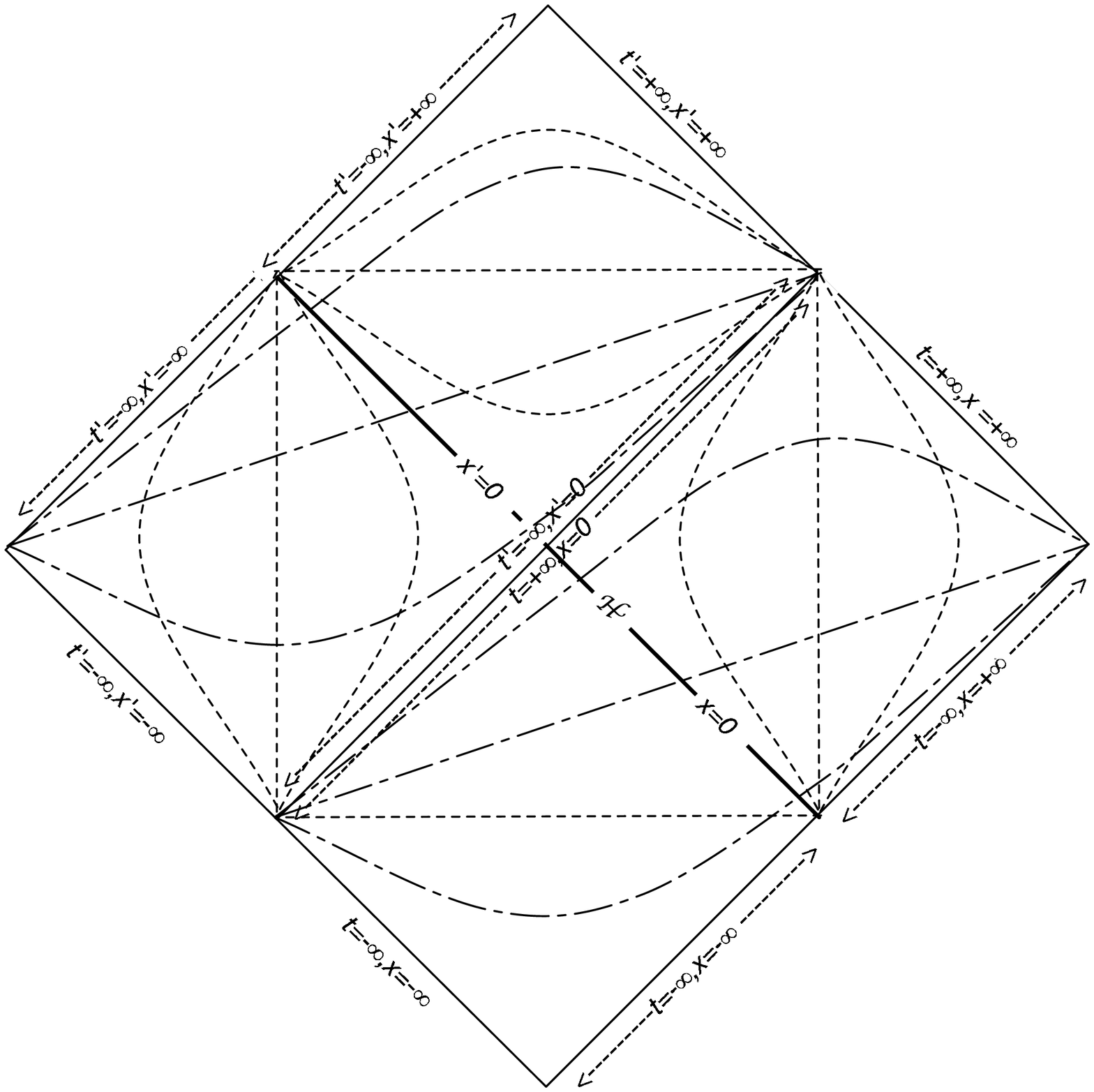}}} \hfil }
\bigskip
\caption{
Conformal diagram of the maximally extended acoustic white hole
geometry.%
} \label{F:conf-wh-ext}
\end{figure}

In general relativity, the standard way to show that a horizon is a
perfectly regular region of spacetime, is to rewrite the metric in
coordinates that are regular at the horizon, as we did at the
beginning of section~\ref{sec:compact-stat}~\cite{MTW,hell,wald}.
With the same line of reasoning we can see that the geometry described
by the metric (\ref{metric}) can be mathematically extended
beyond the range described by the physical coordinates $t$ and
$x$. Look for example at the conformal diagram of the white hole
acoustic spacetime, presented in figure~\ref{F:conf-wh}. In this
particular acoustic geometry, there are geodesics starting from any
event inside the diagram that can reach ${\Im}^{+}_{\mathrm{left}}$
(the upper diagonal line $t=+\infty$, $x=0$), within a finite
lapse of their affine parameter (see the Appendix for a proof), so the
spacetime is geodesically incomplete.  This feature, together with
the regularity of this region, implies that the white hole geometry
can be mathematically extended beyond its open boundary in the future.
If one requires the extension to be made analytically, then it is not
difficult to see that the maximally extended spacetime would be
the one represented by the conformal diagram in
figure~\ref{F:conf-wh-ext}.  In that diagram, the white hole geometry
is supplemented with a black-hole geometry to give rise to an
inextensible metric manifold.  This discussion, or minor variations
of it, applies to most of the acoustic metrics described in this
paper.  Given a conformal diagram, the reader will not find it
difficult to construct its maximal analytic extension.%

Of course, the extended part of the diagram (the black hole part in
our example), does not exist in physical space and time. It is
somewhat like another parallel world. From the point of view of
observers in the laboratory any discussion about this parallel world
does not make much sense.  This is because the connection between the
two worlds happens at an infinite amount of laboratory time in the
future. However, for internal observers (hypothetical observers whose
internal structure is governed by an exchange of virtual phonons, and
who therefore react directly to the acoustic metric as the physical
metric) this connection happens at a finite amount of proper time; and
so, for them it makes perfectly good mathematical sense to ask whether
there is something beyond this boundary.  In other words: If a signal
is sent from the asymptotic region on the right towards the left
(against the flow), observers in the laboratory will say that the
signal never reaches the sonic point at $x=0$.  In contrast, the
internal observers will believe that by swimming against the flow they
should be able to reach $x=0$ in a finite amount of their proper time,
and so they should also be able to experience what happens once they
cross that point.  However, the fact that $x=0$ can be reached in a
finite lapse of proper time, as opposed to the infinite amount of
laboratory time $t$, is not due to an unfortunate choice of the $t$
coordinate, but rather to a dynamical slowing down of the clocks
carried by the internal observers as the sonic point is
approached.\footnote{Consider for example a ``sound clock,''
  constructed by letting a sound pulse bounce between two points
  separated by a constant distance $d$ along a direction orthogonal to
  the flow in the pipe.  For such a clock located at a given value of
  $x$, the round-trip time (as measured in the laboratory) is
  $T(x)=2d/C(x)$, where $C(x)$ is the pulse speed with respect to the
  laboratory. The latter is easily found by requiring that in the
  frame comoving with the fluid the pulse speed be still $c$, so
$C(x)=\left(c^2-v(x)^2\right)^{1/2}$ and%
\[ T(x)=\frac{2d}{c}\left(1-v(x)^2/c^2\right)^{-1/2}\;.\]%
If the internal observers use $T(x)$ as their standard of time, it
is immediate to account for the relation between $t$ and their
proper time, as predicted by the metric (\ref{metric}).}  Hence,
the internal observers will actually be hibernating near the sonic
point, and will remain there forever from our laboratory point
of view.%

There is another important issue of physics hiding in the mathematics
of maximal analytic extension.  While there is no doubt that maximal
analytic extension can successfully be carried out as a mathematical
exercise, one should also think about the physics of the
approximations being made.  We have already made clear that maximal
analytic extension only appears to make sense for hypothetical
internal observers that couple only to the acoustic metric
(\ref{metric}), and do not see the physical spacetime metric of
special relativity.  But this is at best a low-energy approximation;
at high enough energy the atomic nature of matter comes into play and
ultimately phonon dispersion relations turn over~\cite{breaking},
destroying the acoustic approximation.  Indeed the
physically-motivated breakdown of the notion of maximal analytical
extension due to the high-energy breakdown of the relativistic
approximation seems to be a general feature of analog spacetimes.  For
example, this also shows up in analog models based on fermionic
quasi-particles in ${}^3$He-A as noticed
in~\cite{JV,JK}.\footnote{Those authors also consider Carter--Penrose
diagrams for that specific model, obtaining similar but not identical
causal structures.}

In contrast, in standard general relativity with a single unique
physical metric and strict adherence to the Einstein Equivalence
Principle, the absence of such problems related to maximal analytic
extension is automatic. In multi-metric theories, or indeed any theory
that does not impose strict adherence to the Einstein Equivalence
Principle, we may find that the process of maximal analytic extension
fails for physical (rather than mathematical) reasons.

If one takes the view (which is \emph{not\/} standard within the
general relativity community) that Einstein gravity is simply a
low-energy approximation to some radically different and more
fundamental theory of quantum gravity, then this observation based on
the analogue models suggests that there might similarly be physical
problems with maximal analytic extension in real physical spacetimes,
not because there is anything wrong with the mathematics of maximal
analytic extension, but because the physical hypotheses and
approximations underlying the use of differential geometry and the
mathematical machinery of manifolds may break down at the locations
where analytic extension seems mathematically natural.

Within the context of the acoustic geometries, this physical
breakdown in the process of analytic extension is in a sense
obvious --- but we submit that without having a concrete physical
example of this process at hand it would be difficult to see
why the mathematically well-defined process of analytic extension
should be viewed with some circumspection for physics reasons.
In particular, note that the arguments we have used in order
to exclude the possibility, for the internal observers, of
``crossing the boundary at $t=+\infty$,'' rely on the existence of
a privileged external structure --- the laboratory.  Without it,
from the perspective offered by the metric (\ref{metric}) alone,
we could not discard the possibility for the parallel world beyond
$t=+\infty$ to exist.%

\section{Summary and conclusions}%
\label{sec:comments}%
\setcounter{equation}{0}%

Many properties of acoustics in moving fluids can be understood by
recourse to the tools and notions that appear in the study of fields
and rays in effective spacetime geometries. On the other hand, the
analogue gravity
programme~\cite{abh,normal,fischer,bec,silke,silke2,frw,frw2} also
aims to shed light on different problems of the physics of fields in
curved spacetime backgrounds by analyzing acoustic systems (more
generally, condensed matter systems) with similar ``effective''
geometrical properties. In particular, in this paper our interest have
been centered on (1+1)-dimensional background geometries containing
sonic horizons, either present since the beginning of time or with
apparent horizons being created at a certain moment. We have produced
a rather complete catalogue of 1+1 acoustic spacetimes taking into
account their naturalness as models of acoustic effective manifolds in
a condensed matter laboratory.  The specific characteristics of each
of these spacetimes will have important consequences regarding vacuum
polarization effects. We will analyze this issue in a follow-up paper.

We have separated the different geometries into subclasses,
depending on the existence of either one or two sonic
points, and to their eternal or dynamical character. Then, we
have chosen specific velocity profiles (that is, specific 1+1
spacetime geometries), in such a way that they are
easy-to-manipulate representatives of each subclass. Using these
profiles, we have calculated the form of the null coordinates $u$
and $w$ as a function of the laboratory coordinates $t$ and
$x$. In this way, we have described the behaviour of null rays
(geometric acoustics) in each of these geometries.%

In this paper we have been primarily interested in the global
causal structures of these acoustic manifolds. Two metrics
differing only by a conformal factor share the same causal
structure. Therefore, to describe the causal structures of our
catalogue of acoustic spacetimes we have constructed their
conformal diagrams.%

Finally, by looking at the conformal diagram of a white hole spacetime
(we have taken this particular spacetime as a simple and illustrative
example), we have discussed the different views that one could have
regarding the extendibility of that geometry, depending on whether one
views the situation as an internal or an external (laboratory)
observer.  An observer in the laboratory will be reluctant to think
about anything existing beyond its space and time, whereas for a
hypothetical internal observer, whose analysis of the world is
performed only through acoustic experiments, the notion of maximal
analytic extension would seem to make perfectly good sense. However
once we take note of the fact that acoustics in general is only a
low-energy approximation to more fundamental underlying physics, we
see that analytic extension, while perfectly good mathematics, is
physically dubious.  This \emph{might\/} have implications for real
physical gravity if one adopts the perhaps unpopular viewpoint that
what we know as Einstein gravity (and in particular the approximations
underlying the use of manifolds and differential geometry) are only
low-energy manifestations of some radically different
underlying theory.%

\section*{Appendix: Geodesic incompleteness}%
\label{sec:appendix}%
\renewcommand{\theequation}{A.\arabic{equation}}%
\setcounter{equation}{0}%

Let us prove that any acoustic spacetime $({\cal M},\g)$, as
defined in section~\ref{S:fundamental}, which is associated with a
flow that contains sonic points, is geodesically incomplete.  We
shall show this by arguing that the affine parameter along a
geodesic attains a finite value as a sonic point is approached.
This is true for timelike and spacelike geodesics (for which the
affine parameter is simply proportional to proper length along the
curve), as well as for null geodesics.  For the sake of
definiteness, we shall restrict ourselves to a stationary
situation, and we shall focus on the case in which the sonic point
corresponds to the asymptotic line $u=+\infty$. Furthermore, we
shall assume that $x_S=0$. Other cases can be easily dealt with in
a similar way.%

If $\lambda$ is an affine parameter, the geodesic equation is%
\begin{equation}%
\frac{\d^2 x^a}{\d\lambda^2}+{\Gamma^a}_{bc}\,\frac{\d
x^b}{\d\lambda}\,\frac{\d x^c}{\d\lambda}=0\;.%
\label{geod}%
\end{equation}%
It is convenient to work in null coordinates.  Then it will be
enough to consider the equation%
\begin{equation}%
\frac{\d^2 u}{\d\lambda^2}+{\Gamma^u}_{bc}\,\frac{\d
x^b}{\d\lambda}\,\frac{\d x^c}{\d\lambda}=0\;.%
\label{geod-u}%
\end{equation}%
Since the only non-vanishing Christoffel coefficient of the type
${\Gamma^u}_{bc}$ is ${\Gamma^u}_{uu}$, this equation becomes%
\begin{equation}%
\frac{\d^2 u}{\d\lambda^2}
+{\Gamma^u}_{uu}\left(\frac{\d u}{\d\lambda}\right)^2=0\;,%
\label{geodesic}%
\end{equation}%
with%
\begin{equation}%
{\Gamma^u}_{uu}=\g^{uw}\,\frac{\partial\g_{uw}}{\partial u}
=\frac{\partial}{\partial u}\ln|\g_{uw}|=\frac{\partial}{\partial
u}\,\ln\left(\Omega^2\,|c^2-v^2|\right)\;,%
\label{Gamma}%
\end{equation}%
where the metric (\ref{metric-UW}) with $F=G=1$ has been used. For
our purpose, we need only know the asymptotic form of $\lambda(u)$
as $u\to +\infty$.  We can then use equation (\ref{v}) in order to
write, asymptotically,%
\begin{equation}%
|c^2-v^2|\sim 2\,c\,\kappa\,|x|%
\label{|c2-v2|}%
\end{equation}%
and%
\begin{equation}%
{\Gamma^u}_{uu}\sim \frac{\partial\ln|x|}{\partial u}
+\frac{2}{\Omega}\,\frac{\partial\Omega}{\partial u}\;.%
\label{Gamma2}%
\end{equation}%
If we assume that $\Omega$ tends to a finite value at the sonic
point (a very plausible hypothesis, given that it will usually be
some function of density), we can ignore the last term and simply
write%
\begin{equation}%
{\Gamma^u}_{uu}\sim \frac{\partial\ln|x|}{\partial u}\;.%
\label{Gamma3}%
\end{equation}%
Since $\ln|x|\sim \kappa(w-u)$ by equation (\ref{U-asympt}), we find
${\Gamma^u}_{uu}\sim -\kappa$.  Then equation (\ref{geodesic})
implies%
\begin{equation}%
\lambda\sim A-B{\rm e}^{-\kappa u}\;,%
\label{lambda}%
\end{equation}%
with $A$ and $B$ arbitrary constants.  For $u\to +\infty$, the
affine parameter $\lambda$ attains a finite value.%

\section*{Acknowledgements}%

The research of Carlos Barcel\'o is supported by the Education Council
of the Junta de Andaluc\'\i{}a (Spain).  Matt Visser is supported by a
Marsden grant administered by the Royal Society of New Zealand.


%
\end{document}